\documentclass[prx,amsmath,twocolumn]{revtex4-2}
\usepackage[normalem]{ulem}
\usepackage{epsfig}
\usepackage{amsfonts} 
\usepackage{amsmath} 
\usepackage{nicefrac}
\usepackage{color} 
\usepackage{hyperref}
\setcitestyle{square}
\usepackage{subfigure} 
\usepackage{natbib} 
\usepackage{graphicx,amssymb}

\newcommand{\ignore}[1]{}

\usepackage{mathrsfs}
\usepackage{commath}

\usepackage{gensymb} 
\usepackage{mathtools} 
\usepackage[makeroom]{cancel}

\setcitestyle{square}
\usepackage{subfigure}

\usepackage{enumitem}
\usepackage{lipsum}

\usepackage{pbox} % For breaking lines in table
\usepackage{tcolorbox}  % For a gray background textbox

\usepackage{mathrsfs}
\usepackage{commath} 
\usepackage{gensymb} 
\usepackage{mathtools} 
\usepackage[makeroom]{cancel}

\newcommand{\tasd}{\overline{\delta^2(s,t)}}

\newcommand{\average}[1]{\left< #1 \right>}

\begin{document}

\title{Unravelling the origins of anomalous diffusion: from molecules to migrating storks} 
\author{Ohad Vilk$^{1,2}$}
%\email{ohad.vilk@mail.huji.ac.il}
\author{Erez Aghion$^{3}$}
\author{Tal Avgar$^{4}$}
\author{Carsten Beta$^{5}$}
\author{Oliver Nagel$^{5}$}
\author{Adal Sabri$^{6}$}
\author{Raphael Sarfati$^{7}$}
\author{Daniel K. Schwartz$^{7}$}
\author{Matthias Weiss$^{6}$}
\author{Diego Krapf$^{8}$}
\author{Ran Nathan$^{2}$}
\author{Ralf Metzler$^{5}$}\email{rmetzler@uni-potsdam.de}
\author{Michael Assaf$^{1,5}$}
\email{ michael.assaf@mail.huji.ac.il}
\affiliation{$^{1}$Racah Institute of Physics, The Hebrew University of Jerusalem, Jerusalem 91904, Israel}
\affiliation{$^{2}$Movement Ecology Lab, Department of Ecology, Evolution and Behavior, Alexander Silberman Institute of Life Sciences, 
 Faculty of Science, The Hebrew University of Jerusalem, Jerusalem 91904, Israel}
\affiliation{$^{3}$Departments of Physics and Chemistry, University of Massachusetts Boston, MA 02125, USA} 
\affiliation{$^{4}$Wildlife Space-Use Ecology Lab, Department of Wildland Resources and Ecology Center, Utah State University, Logan, UT 84332, USA} 
\affiliation{$^{5}$Institute of Physics and Astronomy, University of Potsdam, Potsdam 14476, Germany} 
\affiliation{$^{6}$Experimental Physics I, University of Bayreuth, D-95440 Bayreuth, Germany}
\affiliation{$^{7}$Department of Chemical and Biological Engineering, University of Colorado Boulder, Boulder, CO 80309, USA}
\affiliation{$^{8}$Department of Electrical and Computer Engineering, and School of Biomedical Engineering, Colorado State University, Fort Collins, CO 80523, USA}
%\email{To whom correspondence should be addressed; E-mail:  michael.assaf@mail.huji.ac.il}

%%%%%%%%%%%%%%%%%%%%%%%%%%%%%%%%%%%%%%%%%%%%%%%%%%%%%%%%%%%%%%%%%%%%%%%%%%%%%%%
%
% A B S T R A C T 
%
%%%%%%%%%%%%%%%%%%%%%%%%%%%%%%%%%%%%%%%%%%%%%%%%%%%%%%%%%%%%%%%%%%%%%%%%%%%%%%%
\begin{abstract}
  Anomalous diffusion or, more generally, anomalous transport, with nonlinear dependence of the mean-squared displacement on the measurement time, is ubiquitous in nature. It has been observed in processes ranging from microscopic movement of molecules to macroscopic, large-scale paths of migrating birds. Using data from multiple empirical systems, spanning 12 orders of magnitude in length and 8 orders of magnitude in time, we employ a method to detect the individual underlying origins of anomalous diffusion and transport in the data. This method decomposes anomalous transport into three primary effects: long-range correlations (``Joseph effect"), fat-tailed  probability density of increments (``Noah effect"), and nonstationarity (``Moses effect"). We show that such a decomposition of real-life data allows to infer nontrivial behavioral predictions, and to resolve open questions in the fields of single particle tracking in living cells and movement ecology.
 
\end{abstract}

\maketitle{}

%%%%%

%%%%%%%%%%%%%%%%%%%%%%%%%%%%%%%%%%%%%%%%%%%%%%%%%%%%%%%%%%%
%%%%%%%%%%%%%%%%%%%%%%%%%%%%%%%%%%%%%%%%%%%%%%%%%%%%%%%%%%%
%%%%%%%%%%%%%%%%%%%%%%%%%%%%%%%%%%%%%%%%%%%%%%%%%%%%%%%%%%%
%%%%%%%%%%%%%%%%%%%%%%%%%%%%%%%%%%%%%%%%%%%%%%%%%%%%%%%%%%%
%%%%%%%%%%%%%%%%%%%%%%%%%%%%%%%%%%%%%%%%%%%%%%%%%%%%%%%%%%% 

\section{Introduction}
Normal diffusion or transport processes obey the Gaussian central limit theorem (CLT) and are ergodic; \textit{i.e.}, mean values of various observables in the system do not depend on the  averaging method. The CLT states that, if a random time series $x(t)$ is the sum of random variables which are (i) identically distributed (with a stationary distribution), (ii) have a finite variance, and (iii) are independent, the probability density function (PDF) $P(x,t)$ of $x$ at time $t$ has a Gaussian shape (\iffalse Appendix~\ref{SubsecGaussianCLT}\fi see Sec. \ref{SecTheoreticalBackground}). The mean-squared displacement (MSD) then satisfies $\langle x^2(t)\rangle \propto t$ at long times, where $\langle\cdot\rangle$ denotes ensemble averaging (EA).   
%While these predictions allow to utilize standard tools to study the system, in practice many natural
Yet, advances in high fidelity methods for single-particle tracking ~\cite{pavani2009three,shen2017single} 
%,wang2021principles, 
and detailed data of animal paths~\cite{humphries2012foraging,
%cooke2016remote,
toledo2020cognitive} show that many  natural processes are in fact \textit{anomalous}, as they violate (some of) the CLT's conditions~\cite{mendez2016stochastic}. %Examples range across many temporal and spatial classical measurement scales. 
 Condition (i) can be violated, \textit{e.g.}, 
%if  energy is dissipated, leading to a slowdown of the process over time, or alternatively 
when the measured
trajectories are confined for increasingly long periods in certain spatial regions, hindering their expansion. Condition (ii) can be violated, \textit{e.g.}, in financial time series, where large fluctuations are highly probable. %\cite{mandelbrot2001scaling}%\cite{mandelbrot2001scaling,mandelbrot2001stochastic} 
%, or e.g., in volcanic-seismic data ~\cite{beccar2020levy}
 %In ecology, this type of violation is often mentioned in the context of the L\'evy flight foraging hypothesis~\cite{viswanathan2011physics, LevyWalks}, more on this below.  
Condition (iii) can be violated, \textit{e.g.}, for biased or correlated motion.  %\erez{[]Ralf's Ref: viscoelastic materials and anti-persistence.]}
Such violations typically yield
\begin{equation} \label{eq0}
    \langle x^2(t)\rangle\propto t^{2H}, 
\end{equation}
with the Hurst exponent being $H\neq1/2$.

Given an empirical time series displaying anomalous transport, the ability to distinguish between the various violations of the CLT is crucial, \textit{e.g.},
%for understanding the basic structure of the system, and 
to determine the system's expansion rate~\cite{LevyWalks,
metzler2014anomalous},  % metzler2019brownian
 rare event statistics ~\cite{kozlowska2005anomalous,assaf2017wkb}
%assaf2010extinction,
%wang2018renewal,
%lawley2020extreme
and method of averaging~\cite{bouchaud1992weak,
%rebenshtok2007distribution,
burov2011single,
%manzo2015weak
vilk2021ergodicity}, as well as to infer features in the diffusion medium~\cite{szymanski2009elucidating,massignan2014nonergodic,krapf2015mechanisms,cherstvy2019non} and elucidate the underlying microscopic process.  
However, this characterization remains a major challenge in various fields including single particle tracking and movement ecology~\cite{nathan2008movement,chakraborty2019nanoparticle,janczura2021identifying},  and much effort is made to  develop techniques to tackle it; see, \textit{e.g.}, Refs.~\cite{kepten2013improved, fleming2014fine, sabri2020elucidating,thapa2018bayesian}. %\red{[[Added Ref.~\cite{thapa2018bayesian}.]]}
% [[Originally was:]] ~\cite{%kepten2013improved
% munoz2019machine,
%fleming2014fine,
%bo2019measurement, 
%sabri2020elucidating, thapa2021leveraging 
%capuani2020mini,
%janczura2020machine,munoz2020andi,
%argun2021classification,
%ANDI2021
% \cite{krog2018bayesian,park2021bayesian,thapa2022bayesian}}thapa2018bayesian }
Recently, machine-learning methods for analyzing anomalous transport data have been widely studied, see e.g.,~\cite{loch2020impact, kowalek2019classification}, 
%munoz2020single,manzo2021extreme,gentili2021characterization,argun2021classification,Dezhong2021,pineda2022geometric, 
and for many applications they were shown to outperform estimators based on classical statistics~\cite{ANDI2021}.  Yet, the ``black box" nature of these data-driven algorithms may  hinder the ability to account for the underlying reasons of the observed phenomena \cite{ANDI2021}.

Here, based on positional (tracking) data, we employ a specialized three-effect decomposition method~\cite{mandelbrot1968noah,chen2017anomalous} to disentangle the effects leading to anomalous transport,  without making prior assumptions on the underlying model governing the dynamics. By analysing three independent  properties of the time series  presented below, we  determine whether the measured diffusion is anomalous due to violation of condition (i), (ii) and/or (iii) above. 
To establish the broad applicability of the technique, we study empirical data sets that range over 12 orders of magnitude in length ($10^{-6}$--$10^6$ m), and 8 orders of magnitude in time ($10^{-3}$--$10^5$ s). %Beyond empirical data,  
We also present results of numerical simulations of random walk models, which have been previously proposed to describe some of these systems. Notably, applying this method provides important insight into key open questions in various scientific fields, as detailed below. Thus, we aim at promoting this method as a common practice for future empirical studies of anomalous transport. 
%For more details on the data sets, and the results of the analysis of all the different systems, see  Sec.~\ref{SecMainAnalysis}. For details on the simulated systems see Methods.

%\begin{figure*}[t]
%\centering
%\includegraphics[width=1\linewidth]{Picture1AnalysisScheme125.png}
 %\caption {\footnotesize{5}}
 %\label{fig:Scheme}
%\end{figure*} 

\begin{figure*}[t]
\centering
\includegraphics[width=1\linewidth]{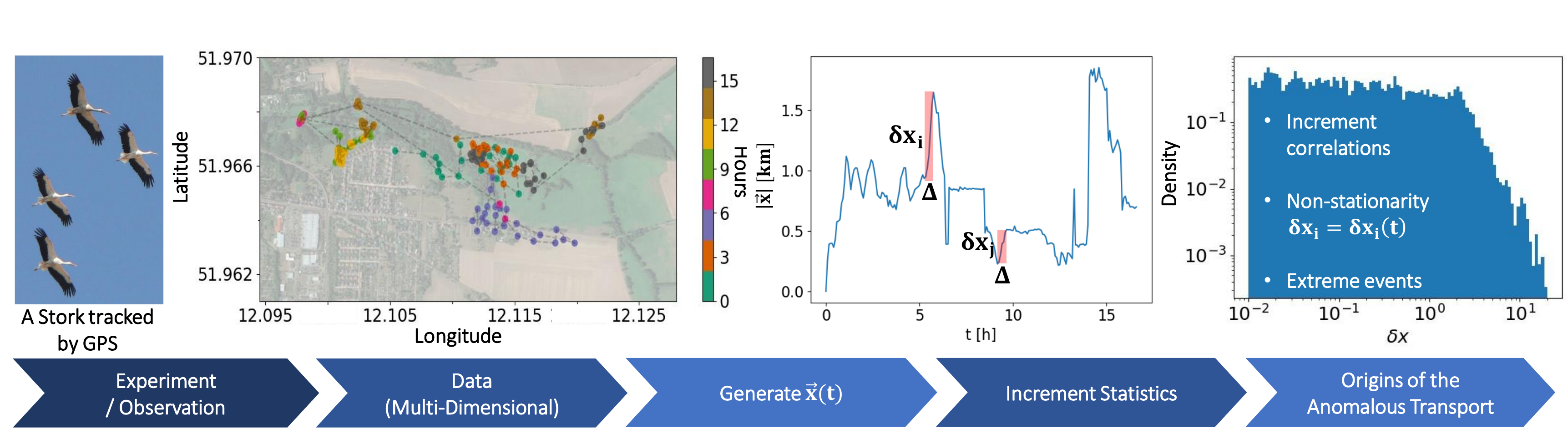} 
 \vspace{-5mm}
 \caption {\footnotesize{Scheme of the three-effect decomposition of the origins of anomalous transport measured in data. (left to right) (i) Collect data from an experiment. (ii) Generate multidimensional paths from experimental observations (\textit{e.g.}, a flight pattern of a stork tracked by GPS over several hours). (iii) Generate the time series $\vec{x}(t)\equiv\mathbf{x}(t)$, and decompose it into vector increments $\delta \mathbf{x}_i=\mathbf{x}_{i\Delta}-\mathbf{x}_{(i-1)\Delta}$ ($i=1, ..., N$) of equal duration $0<\Delta\ll t$. 
 %Note that on the second panel from the right we only present the time-series of the displacement norm $|\mathbf{x}(t)|$, although the decomposition to increments is done for $\mathbf{x}(t)$ and only later we calculate the displacement norm. 
 (iv) Obtain the statistics of increment sizes. (v) Determine whether the process is correlated, explicitly time dependent, or prone to extreme fluctuations due to a fat-tailed velocity PDF. 
 }} 
 \label{fig:Scheme}
\end{figure*} 

%%%%% 
%%%%%\
%%%%% 
%%%%% 

In our analysis below, we study the empirical data from the various experiments as a stochastic process of the form  $\mathbf{x}(t)$, where $t$ is the measurement time, in $d\geq1$ dimensions (vectors are denoted in bold font). For instance, $\mathbf{x}(t)$ can represent  a time-series of the distance traveled by a bird from its nest in the course of one day, as function of time, where one can always set $\mathbf{x}(0)=\mathbf{0}$. The process $\mathbf{x}(t)$ can be described by a discrete sum of random increments, $\mathbf{x}(t)=\sum_{j=1}^N \delta \mathbf{x}_j$, where $\delta \mathbf{x}_j \!\equiv\! \mathbf{x}(j\Delta) \!-\! \mathbf{x}([j\!-\!1]\Delta)$ and $N\!=\!t/\Delta$, while $0\!<\!\Delta\!\ll\! t$ is an arbitrary time increment. Moreover,  $\mathbf{v}_j\equiv \delta \mathbf{x}_j/\Delta$ is the average velocity vector in the $j$th increment, and the \textit{velocity} PDF, $P(|\mathbf{v}|,t)$,  is the probability density of its absolute value. %When we observe an ensemble of $\mathbf{x}(t)$s at increasing times,  if we find it to be anomalous, e.g., in the sense that its average squared value grows more quickly or less quickly than the normal (linear) rate expected by the CLT - we wish to understand \textit{why} that is the case; ``Is the distance traveled by the animal (for example) growing slowly in time since it is increasingly crossing through harder-than-normal terrain during its path?", ``Is it just an unusual statistical fluctuation, where the animal felt the presence of a predator in the area and decided to hide for a longer-than-usual time sometime during the day"? Or maybe; "Is the animal deliberately avoiding long trips today, and wants to stay close to its nest"? The questions are similar when we talk about other anomalous processes, such as rain records: "are they gradually dropping with time?", are they ``affected by a random singular drought event?" or ``is the lack of rain in the recent weeks, leading to dwindling water sources and \textit{consequently} leads to less rain in the future?". 
To distinguish between the above three different ways of violating the CLT in an empirical time series, we compute the corresponding increments of $\mathbf{x}(t)$  and analyze their size-statistics~\cite{mandelbrot1968noah,chen2017anomalous}, as shown in Fig.~\ref{fig:Scheme}. We compute three  numerical values that describe the temporal scaling of three observables: (i)  mean absolute velocity $\average{|\mathbf{v}|}$, (ii) mean-squared velocity $\average{\mathbf{v}^2}$, and (iii) ensemble-averaged time-averaged MSD (TAMSD) $\langle\overline{\delta^2(s, t)}\rangle$, see below~\cite{aghion2021moses}. 
% This is explained below, and the Supplementary, Table S1. 

The manuscript is organized as follows.
In Sec.~\ref{MaterialsAndMethods} and \ref{SecTheoreticalBackground}, we detail our empirical setups and the mathematical background for the three-effect decomposition method, respectively. In Sec. \ref{SecMainAnalysis}, we provide the main results of our analysis of the empirical setups. Finally, in Sec. \ref{SecDiscussion} we  provide a data-based discussion on the relations between different violations of the CLT in our empirical setups, and discuss additional methods that may extend the analysis in future research.

\section{Empirical setups}
\label{MaterialsAndMethods}

We detect the origins of anomalous diffusion by employing the three-effect decomposition method in 7 empirical systems, comprised of 16 empirical setups. Here, we provide technical details for all systems, which are organized by ascending physical size and temporal range. For statistical analysis and results see Sec.~\ref{SecMainAnalysis}.

\subsection{Rhodamine molecules}
A solution of rhodamine 6G molecules was deposited onto a cleaned borosilicate glass coverslip, then dried in a vacuum chamber for 30 min. The dry surface was then exposed to various degrees of ambient relative humidity between 30\% and 100\%, which resulted in the equilibrium condensation of water nanofilms of a few (1-8) molecular layers, with thicknesses that increased systematically with increasing humidity. %~\cite{sumner2004nature}. 
Individual rhodamine molecules were traced and recorded using a total internal reflection fluorescence (TIRF) microscope (532 nm laser excitation) with image acquisition times of 50 ms. Tracking (object localization and trajectory linking) was performed using  MATLAB code. Roughly $10^4$ trajectories were captured for each condition; see\cite{sarfati2020temporally} for details.

\subsection{Tracer particles in the cytoplasm of mammalian cells}
Tracer particles (Qdot 655 ITK Carboxyl core (CdSe)-shell (ZnS), ThermoFisher, Waltham, MA) were incorporated into HeLa (human cervical cancer) cells by bead loading, followed by a relaxation time of 1~h before imaging. In preparation for this procedure, cells were plated 36-48~h prior to bead loading on $35$~mm diameter $\Delta$T dishes (Bioptech, Butler, PA) for temperature control, coated with 0.5\% matrigel matrix (Corning Life Sciences, NY) for improved adhesion. Depolymerization of actin filaments was induced by adding 200~nM latrunculin~A to the medium directly after bead loading. Images were acquired with an EMCCD camera at 10~frames/s on a custom-built microscope equipped with an Olympus PlanApo $100\times$ NA$1.45$ objective, a CRISP ASI autofocus system, and a MicAO 3DSR adaptive optics system (Imagine Optic, Orsay, France) to correct optical aberrations. Quantum dots were excited at $561$~nm under continuous illumination and trajectories were extracted from image stacks with FIJI/TrackMate. Removing immobile tracks, this approach eventually yielded large data sets from which a random selection of $M=1000$ ($M=200$) tracks with $N=100$ ($N=500$) positions were used for untreated (latrunculin-treated) cells.
For further details, see supplemental material of~\cite{sabri2020elucidating}.

\subsection{Motile amoeba}
Tracking of motile cells was performed with the social amoeba {\it Dictyostelium discoideum}, using AX2 wild-type cells that were cultivated in HL5 medium on polystyrene Petri dishes or in shaken suspension ~\cite{cherstvy2018non}. Prior to imaging, cells were washed, the HL5 medium was renewed, and cells were placed on a plastic Petri dish at an average density between 1 and 2 $\times 10^4$ cells per cm$^2$ and allowed to attach to the surface for 30 minutes. Cells were then recorded for 6 hours with a bright-field microscope at a frame rate of 0.05 Hz. To track the cells, images were segmented and the centers of mass of regions corresponding to the cells were calculated and connected from one frame to the next by nearest neighbor particle tracking. Segmentation and tracking were performed with a custom-made MATLAB algorithm (MathWorks, Ismaning, Germany) using well-established particle-tracking methods. %Afterwards, trajectories were analyzed with Mathematica (Wolfram Research Inc., Champaign, USA). 
If cells were lost during the tracking procedure because they left the field of view, collided with a neighboring cell, or divided, the recorded track ended, and a new trajectory was started once a new cell could be detected in the field of view. Only trajectories with over 60 time steps were used in the analysis; See Ref.~\cite{cherstvy2018non} for details.

\subsection{Harvester ants}
Movement paths of individual harvester ants (\textit{Messor arenarius}; a solitary foraging species) were mapped in 2005 as part of research conducted in "Sayeret-Shaked" park, North-Western Negev desert, Israel (see~\cite{avgar2008linking} for further details).
Individual ants, each from a different colony, were marked with colored fluorescent powder and then tracked by placing numbered pins at their positions every 10 s (with minimal interference to the ant's behavior). Route mapping started once the ant departed the nest (after entering it at least once since being marked) and ended after two consecutive foraging trips (whether successful or not). Pins were then mapped at 1 cm resolution using measurement tapes and a costume-built wooden frame, and the positional time-series was digitized.

\subsection{Black winged kite}
An individual black-winged kite (\textit{Elanus caeruleus}), residing in the Hula Valley, Israel, was tracked using ATLAS, an innovative reverse-GPS system. ATLAS localizes extremely light-weight, low-cost tags~\cite{toledo2020cognitive, vilk2021ergodicity, vilk2022phase}, where each tag transmits a distinct radio signal which is detected by a network of base-stations distributed in the study area. Tag localization is computed using nanosec-scale differences in signal time-of-arrival to each station, alleviating the need to retrieve tags or have power-consuming remote-download capabilities. The kite was tracked for 164 consecutive days in the years 2019-2020, with a mostly constant tracking frequency of 0.25 Hz. 

As in Ref.~\cite{vilk2021ergodicity}, the kite's tracks are segmented into two behavioral modes, local searches (area restricted search) and commuting (directed flights between local searches). Localizations were segmented by detecting switching points in the data -- distinct points in which the bird switches between the two behaviors~\cite{benhamou2014scales}. Switching points were detected using spatiotemporal criteria segmentation, such that localizations that are in proximity to one another, both in space and time were segmented together. %~\cite{gurarie2016animal}. 
In accordance with the conclusions of Ref.~\cite{vilk2021ergodicity} we independently analyze the time series ensembles representing instances of local searches and commuting flights.

\subsection{White stork}
An adult white stork (\textit{Ciconia ciconia}) was tracked between May 2012 and July 2020 with high-resolution GPS, see Ref.~\cite{rotics2016challenges} and Sec. \ref{sec:vulturetech} for more details. Here the GPS location and speed were recorded at a frequency of 1/300 Hz when solar recharge was high (92\% of the time), and otherwise every 20 min. 
We omit days with lower frequency ($<1\%$ of tracked days) and only include localizations that occur after the first recorded velocity of $>4$ m/s (see below). 

\subsection{Eurasian griffon vulture} \label{sec:vulturetech}
An Eurasian griffon vulture (\textit{Gyps fulvus}, Hablizl 1783) was tracked in Israel and surrounding countries with high-resolution global positioning system (GPS), between October 2012 and October 2015. The 90-g GPS transmitters (E-Obs GmbH; Munich, Germany) were fitted in a backpack configuration and set to a 13 h duty cycle, between 7:00 a.m. and 8:00 p.m. to correspond with the vulture's activity pattern~\cite{harel2016decision}. Localizations were optimally recorded at a frequency of 1/600 Hz (73\% of the time), or 1/1200 Hz (23 \% of the time). Vulture days tracked at lower frequencies were omitted from this study. %Transmitting efforts were approved by the Israel Nature and Parks Authority and were in accordance with the ethics guidelines of the Hebrew University of Israel (NS-07-11063-2). 
See Ref.~\cite{harel2016decision} for more details.

As the time of the vulture's departure from the nest can drastically vary between different days, we only include localizations that occur after the first recorded velocity of $>4$ m/s (as done for the stork)~\cite{harel2016decision}.

\section{Theoretical fundamentals} 
\label{SecTheoreticalBackground}
In normal transport processes the first absolute moment satisfies $\langle |\mathbf{x}(t)|\rangle\propto\sqrt{\langle \mathbf{x}^2(t)\rangle}$, such that both observables provide the same information on the process. Thus, for these processes, the MSD [Eq.~(\ref{eq0})] is a direct measure of typical fluctuations. Yet, in anomalous processes, the MSD may diverge, making the Hurst exponent inappropriate for characterizing the dynamics. In other cases, %e.g., for processes exhibiting \textit{strong anomalous diffusion}, 
the scaling with time of $\langle|\mathbf{x}(t)|\rangle$  and MSD,  representing typical and large fluctuations respectively, is different. Consider, e.g., L\'evy walks~\cite{rebenshtok2014non,rebenshtok2016complementary,dentz2015scaling,aghion2018asymptotic}, in one dimension; here the random walker, starting at $x=0$, moves with constant speed $\pm|v|$ (to the right or left) in a series of independent motion intervals, where  the random  interval \textit{duration}, $\tau$, is  power-law distributed $\propto\tau^{-1-\alpha}$ ($1\leq\alpha\leq2$), with  diverging $\langle\tau^2\rangle$. As a result, while most intervals are short, a small fraction are very long, with a corresponding very large travelling distance~\cite{rebenshtok2014non}. Thus, out of a group of walkers that were released simultaneously at the origin, most of the walkers will remain close to $x=0$, while a few may be found very far away. This results in $\langle|\mathbf{x}(t)|\rangle\propto t^{1/\alpha}$, representing the expansion of the bulk of the walker group (PDF width), whereas the MSD scales as $\langle \mathbf{x}^2(t)\rangle\propto t^{3-\alpha}$, as it is strongly affected by rare longest-travelling individuals (PDF tails)~\cite{rebenshtok2014non}. 

\begin{figure*}[t]
\centering
\hspace{-12mm}\includegraphics[width=.9\linewidth]{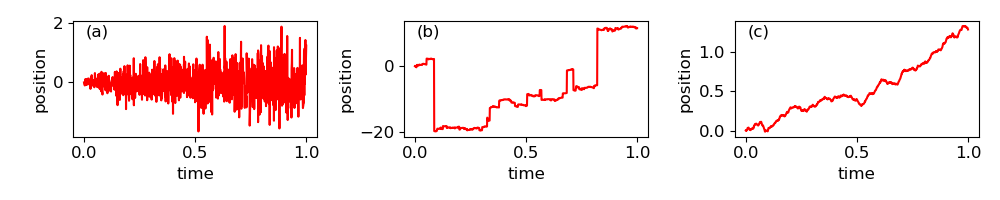}
\includegraphics[width=0.9\linewidth]{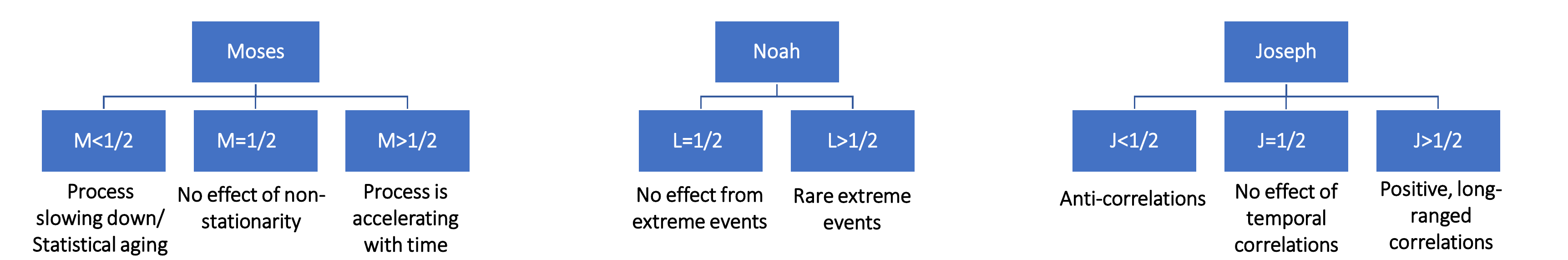}
 \vspace{-2mm}
 \caption {\footnotesize{Prototypical examples of processes displaying anomalous diffusion. (a) A non-stationary accelerating process (scaled Brownian motion), leading to the \textit{Moses} effect. This can result, \textit{e.g.}, from movement in a (temporal) temperature gradient  increasing fluctuations over time. (b) A process with rare extreme events (L\'evy flight), leading to the \textit{Noah} effect. Such a process has been used, \textit{e.g.}, to model flight patterns of the wandering albatross~\cite{viswanathan2011physics}, but this characterization is under debate~\cite{benhamou2007many}. (c) A process with long ranged temporal correlations (fractional Brownian motion), giving rise to a trended motion, and the \textit{Joseph} effect. This may result, \textit{e.g.}, from long-range memory in network traffic~\cite{paxson1995wide, eliazar2013fractional}. 
 In all three panels, due to different violations of the CLT the rate of diffusion is different from normal diffusion. In addition, position and time are shown in arbitrary units.  
 Lower panels: The physical interpretation of the values of the Moses, Noah and Joseph exponents. (i) A Moses effect is a proxy for nonstationarity. Here $M>1/2$ and $M<1/2$ respectively indicate an accelerating and decelerating process (the latter entails aging), whereas for $M=1/2$ the process is stationary. (ii) A Noah effect is a proxy for detecting extreme rare events. Here $L>1/2$ indicates susceptibility to large fluctuations due to a fat-tailed velocity PDF, whereas for $L=1/2$ no Noah effect occurs. (iii) A Joseph effect is a proxy for long-range correlations or anti- correlations. Here $J>1/2$ indicates long-range positive temporal correlations which may lead to superdiffusion, and $J<1/2$ indicates anti-correlations which may lead to subdiffusion. When $J=1/2$ no Joseph effect occurs.}} 
 \label{fig:TimeSeriesExample}
\end{figure*}

The decomposition method derived in~\cite{mandelbrot1968noah,chen2017anomalous} allows one to fully account for such issues by describing anomalous diffusion with three exponents instead of a single one, relying in part on the above distinction between typical and large fluctuations. The method also  enables one to directly obtain the Hurst exponent from  a small ensemble of time series \cite{Meyer2021Decomposing,ANDI2021}. Below, and in Appendix \ref{SubsecGaussianCLT}, we present the definition of the three exponents in the context of the individual type of CLT violation that they quantify, and explain how each violation leads to anomalous diffusion, see also Fig. \ref{fig:TimeSeriesExample} \cite{viswanathan2011physics, benhamou2007many, paxson1995wide, eliazar2013fractional}.

\subsection{The effect of nonstationarity}   
As shown in Appendix \ref{SubsecGaussianCLT}, the first condition for the validity of the CLT is that the process increments are identically distributed. This stems from the condition that the dynamics need to be  statistically invariant at each step of the process.  Violation of the CLT due to increment nonstationarity leading to anomalous diffusion is called the ``\textbf{Moses effect}''~\cite{chen2017anomalous,meyer2018anomalous,
%bormashenko2019moses,
aghion2021moses}.  Here  nonstationarity is quantified by the exponent $M$~\cite{chen2017anomalous}  measured via the scaling of the absolute mean of the velocity vector:
\begin{equation}
    \average{\overline{|\mathbf{v}|(t)}} =\average{\frac{\Delta}{t}\sum_{j = 1}^{t/\Delta} |\mathbf{v}_j|} \propto t^{M - 1/2},
\label{eq1}
\end{equation}
where the overline denotes time averaging (TA).
If the process has stationary increment distribution, $M=1/2$. The Moses effect occurs when $M\neq1/2$, implying either accelerating ($M>1/2$) or decelerating ($M<1/2$) dynamics, \textit{e.g.}, due to aging; the latter means that the process is slowing down with time and thus, the observed dynamics may seem different depending on the measurement time.  
A key consequence of this effect is \textit{weak-ergodicity breaking}~\cite{bouchaud1992weak,bel2005weak,burov2011single}, as ergodicity requires stationarity. In Fig.~\ref{fig:TimeSeriesExample}a we plot an example of a simulated non-stationary accelerating process; below this panel we list the different regimes of $M$ and their physical interpretation. 
Notably, in analogy with the above distinction between the mean absolute-position $\langle|\mathbf{x}(t)|\rangle$ and the MSD, $\langle|\mathbf{v}|\rangle$ is representative of typical increment fluctuations (PDF bulk), described by the diffusion coefficient. Thus, the exponent $M$ effectively represents the nonstationarity in the diffusion coefficient, as observed for the prototypical scaled Brownian motion (SBM) process.

\subsection{Extreme events} 
The second  condition for the validity of the CLT is that the increment variance is finite. As stated above, for some processes displaying anomalous diffusion, the increment variance (similarly as the MSD) is dominated by the tails of the velocity PDF. If the increment variance is infinite or grows in time, condition (ii) of the CLT is violated, which may lead to faster than linear growth of the MSD (Appendix \ref{SubsecGaussianCLT}). 
This effect leading to anomalous diffusion is called the ``\textbf{Noah effect}'' ~\cite{mandelbrot1968noah}. In accordance with  Eq.~\eqref{eq1}, it is quantified by the latent exponent $L$~\footnote{Note that, since the ensemble averaging and summation are commutative, $\average{\overline{\mathbf{v}^2(t)}}$ in Eq.~(\ref{eq2}) can also be written as the sum of mean-squared velocities $=\frac{\Delta}{t}\sum_{j = 1}^{t/\Delta} \langle\mathbf{v}^2_j \rangle $.}:  %the scaling of the %finite-ensemble 
%sample mean 
\begin{equation}
    \average{ \overline{\mathbf{v}^2(t)}} = \average{\frac{\Delta}{t}\sum_{j = 1}^{t/\Delta} \mathbf{v}^2_j } \propto t^{2L + 2M - 2}.
    \label{eq2}
\end{equation}
By definition  $L\geq 1/2$~\cite{mandelbrot1968noah}. If, for a stationary velocity PDF with $M=1/2$, in addition $L=1/2$, $\langle\overline{\mathbf{v}^2(t)}\rangle$ is constant. Yet, if $L>1/2$, its value will grow in time, even though $M=1/2$. In the absence of extreme events $\langle\overline{\mathbf{v}^2(t)}\rangle\propto t^{2M-1}$, which is asymptotically equivalent to  $\langle\overline{|\mathbf{v}(t)|}\rangle^2$, as occurs in Gaussian processes.  
A deviation from this scaling, quantified by $L$, is a proxy for detecting extreme (non-Gaussian) rare events, dominating the path with increasing probability as time evolves.  This occurs, \textit{e.g.}, in L\'evy flights~\cite{LevyWalks,chen2017anomalous}, where the noise is scale-free. An example for a process with rare extreme events is given in Fig.~\ref{fig:TimeSeriesExample}b, where below this panel we list the different regimes of $L$ and their physical interpretation.

\subsection{Temporal autocorrelations} 
Long-ranged temporal autocorrelations and anti-correlations may lead to a non-linear scaling of the MSD, via violation of condition (iii) of the CLT. This is called the ``\textbf{Joseph effect}"~\cite{mandelbrot1968noah,chen2017anomalous, paxson1995wide, eliazar2013fractional},
quantified by the exponent $0\!<\!J\!\leq\!1$, and formally defined via the velocity autocorrelation function (Appendix~\ref{SubsecJAndTASD}) with a positive or negative sign, describing a positively correlated (persistent) or anti-correlated (anti-persistent) process, respectively. 
While $J$ can be measured in various ways~\cite{%Hoell19,
chen2017anomalous,meyer2018anomalous,aghion2021moses}, here we use the TAMSD (Appendix~\ref{SubsecJAndTASD}):
\begin{equation}
\small 
\hspace{-1mm}\average{\tasd} \!=\!{} \left\langle{}\!\! \frac{1}{t\!-\!s}\! \int_0 ^{t\!-\!s}\!\!\!\!\!\!\!
 \left[ \mathbf{x}(t'\!+\!s)\! -\! \mathbf{x}(t')\right]^2  \!\mbox{d} t'\!\!\right\rangle{}\!\!\propto\! t^{2L+2M-2}s^{2J}\!.
 \label{eq4}
\normalsize 
\end{equation} 
For long-ranged temporal correlations (decaying very slowly in time) one has $J\neq1/2$, thus violating the  CLT, which is valid only for short-ranged temporal correlations. Another option is temporal  anti-correlations~\cite{mandelbrot1968noah}. %A violation of the CLT in the case of $J>1/2$, which may lead to superdiffusion, or in the anti-correlated case with $J<1/2$, which may lead to subdiffusion, gives rise to the ``\textbf{Joseph effect}''~\cite{mandelbrot1968noah}.
The driving mechanism behind this effect can be, \textit{e.g.}, biased movement, diffusion in confined space, or long-range memory. An example for a process with long-range memory is given in Fig.~\ref{fig:TimeSeriesExample}c, where below this panel we list the different regimes of $J$ and their physical interpretation. 
%, and the possible scaling of all three exponents is summarized in the lower panel of Fig.~\ref{fig:TimeSeriesExample}. 

\subsection{Connection between the exponents%, and the scope of validity of the method
}
The above definitions  yield a fundamental \textbf{summation relation} between $M$, $L$, $J$ and $H$~\cite{chen2017anomalous,meyer2018anomalous,aghion2021moses}:
\begin{equation} \label{summation_relation}
    H = J + L + M - 1,  
\end{equation} 
This relation, connecting these three effects, is  central for all the results presented below, and is confirmed by analysing a large variety of empirical systems~\footnote{From Eqs.~(\ref{eq0})-(\ref{eq4}) it is evident that an observed process resembles simple Brownian motion, if $L=M=J=H=1/2$.}. The summation relation [Eq.~\eqref{summation_relation}] is derived analytically for $J>1/2$ using the Green-Kubo relation~\cite{meyer2018anomalous,aghion2017large}, whereas for $J<1/2$,  it is derived directly from the autocorrelation function of fractional Gaussian noise~\cite{aghion2021moses}, which is commonly used in  modeling of processes with long-ranged anti-correlations, see e.g.,~\cite{Beran}. Importantly, it can be shown that the validity of the three-effect decomposition method  and the resulting summation rule hold for any process that satisfies (i) the power-law scaling of Eqs.~\eqref{eq1} and (\ref{eq2}), at least locally over some finite time interval, and (ii) $\int_0^s\langle\mathbf{v}(t)\mathbf{v}(t+\tilde{s})\rangle d\tilde{s}\propto s^{2J-1}$ at large $s$ (see Appendix~\ref{SubsecJAndTASD} and~\cite{meyer2017greenkubo,dechant2014scaling,aghion2021moses})~\footnote{Equation~\eqref{summation_relation} was shown to be valid for a wide variety of stochastic models such as \textit{scaled Brownian motion}~\cite{lim2002self,chen2017anomalous}, L\'evy flight~\cite{mandelbrot1968noah}, FBM and ARFIMA processes~\cite{Man68fbm,aghion2021moses}, annealed transient time random motion~\cite{massignan2014nonergodic}, CTRW~\cite{scher1975anomalous}, L\'evy walks~\cite{LevyWalks}, and deterministic weakly chaotic maps~\cite{meyer2018anomalous}.
%Moreover, it was shown to be valid for other standard processes~\cite{chen2017anomalous,meyer2018anomalous,ANDI2021,aghion2021moses,Meyer2021Decomposing,fox2021aging}, such as 
%In was  proposed~\cite{chen2017anomalous} that its validity of thi. 
In contrast, Eq.~(\ref{summation_relation}) breaks down for multiplicative processes, such as geometric Brownian motion~\cite{peters2013ergodicity}, with an exponentially-growing mean. It also cannot be applied to ultraslow diffusion processes, where the MSD grows logarithmically in time, e.g., in Sinai diffusion~\cite{sinai1983limiting}, granular gases in the homogeneous cooling regime~\cite{bodrova2016underdamped}, ultraslow CTRW processes~\cite{havlin1990new} and ageing-CTRW~\cite{lomholt2013microscopic}.}. 
Note that, the summation rule means that the above three effects are exhaustive for violating the CLT.

\section{Results}
\label{SecMainAnalysis} 

% \michael{\textbf{Ohad - for coherence, please check that each time M<1/2 you write negative effect, and same with L and J.}}

% We detect the origins of anomalous diffusion in the 7 systems detailed above, which are comprised of 16 empirical setups. : rhodamine molecules diffusing within water nanofilms %(thickness: 1-8 H$_2$O layers) adsorbed at the silica-air interface, 
% with varying relative humidity~\cite{sarfati2020temporally};  tracer particles (quantum dots) in the cytoplasm of untreated and treated mammalian cells~\cite{sabri2020elucidating}; flat amoeba cells ({\it Dictyostelium discoideum}) diffusing on a surface~\cite{cherstvy2018non}; harvester ants (\textit{Messor arenarius}) embarking on foraging trips~\cite{avgar2008linking}; a single black-winged kite (\textit{Elanus caeruleus}) during commuting and area-restricted searches~\cite{vilk2021ergodicity}; a single white stork (\textit{Ciconia ciconia})~\cite{rotics2016challenges} clustered according to seasonality;  and a single Eurasian griffon vulture (\textit{Gyps fulvus}, Hablizl 1783)~\cite{harel2016decision}. 
% %For details on all systems and experiments see Methods. 
% Each  
% %empirical 
% setup is represented by an ensemble of time series, for which we obtain statistics in terms of the above quantities. For details on the experimental setups, mathematical derivations and error analysis, see Methods (below) and Appen.~\ref{AppenSec1_3}\iffalse Supplemental Sec. S1.3\fi.

For each setup described in Sec.~\ref{MaterialsAndMethods}, represented by an ensemble of time series,  we obtain statistics in terms of the quantities given by Eqs.~(\ref{eq0}-\ref{eq4}); for details on the statistical analysis see Appendix~\ref{SubSecLocalScaling}.
In Table~\ref{tab:exponents}, we summarize the scaling exponents $J, L, M$, measured for all experimental systems, along with the predicted value of $H$ based on Eq.~\eqref{summation_relation}, denoted by $H_p$.
Remarkably, for all data sets we find good agreement between $H$ determined from $\average{x^2(t)}$ and $H_p$, with a relative error $\leq$ 10\%, thus confirming the validity of Eq.~\eqref{summation_relation} in the empirical data. 
In most of the studied data sets anomalous diffusion is primarily caused by the \textit{Joseph and Moses effects}; the \textit{Noah effect} was only observed for the stork and searching kite.
We now list the exponents found for each empirical setup, suggest \textit{plausible} models, and discuss various implications of our findings. Below we present figures for three prototypical examples: amoebas, stork and vulture; for the rest see Supplemental Material (SM)~\cite{SM}, Sec. S1, Figs. S1-S11. %particular data set ordered according to the system's typical length scale, from diffusing molecules to migrating storks. 

%%%%%%%%%%%%%%%%%%%%%%%%%%%%%%%%%%%%%%%%%%%% Table: 
%%%%%%%%%%%%%%%%%%%%%%%%%%%%%%%%%%%%%%%

%\newgeometry{width=20.6cm, left=0.4cm, top = 0.5cm}
 %%%%%%%%% Start Table 
%\begin{widetext} 
%\begin{center} 
\begin{table*}[t!] 
\begin{tcolorbox}[colback=white!10]
\footnotesize
\centering 
\begin{tabular}{p{2.5cm}p{2cm}p{2.5cm}p{1cm}p{1cm}p{1cm}p{2cm}p{2cm}p{1.2cm}p{0.1cm}} %ll c ccccc
\hline\hline 
dataset        & ensemble size & regime      & J    & L    & M    & H  measured  & H prediction & $\frac{|H_p - H|}{H}$ \\
[0.5ex]
\hline
Rhodamine   100\%  &  174  & $50 < t < 1500$ ms & 0.50 & 0.50  & 0.41 & 0.38 $\pm$ 0.02 & 0.42 $\pm$ 0.04 &  10\% \\
Rhodamine 90\% & 298 &  $50 < t < 1500$ ms & 0.38 & 0.50  & 0.42 & 0.28 $\pm$ 0.02 & 0.30 $\pm$ 0.03 &  7\%  \\
Rhodamine 85\%  &  239 & $50 < t < 1500$ ms & 0.34 & 0.50 & 0.40 & 0.22 $\pm$ 0.02 & 0.24 $\pm$ 0.05 & 9\%    \\
Rhodamine 75\%  & 258 & $50 < t < 1500$ ms & 0.22 & 0.51  & 0.44 & 0.18 $\pm$ 0.02 & 0.18 $\pm$ 0.01 &  $<$1\% \\
Rhodamine   30\%  & 436 & $50 < t < 1500$ ms & 0.09  & 0.50 & 0.44 & 0.07 $\pm$ 0.02 & 0.04 $\pm$ 0.03 &  ** \\ 
Tracers in  &  200 & 0.1 $< t <$ 5 s & 0.39 & 0.50 & 0.41 & 0.31 $\pm$ 0.01 & 0.30 $\pm$ 0.02        & 3\%           \\
\;\;\;treated cells     &    & 5 $ <t <$ 50  s  & 0.50 & 0.50 & 0.44 & 0.44 $\pm$ 0.01 & 0.44 $\pm$ 0.01        & $<$1\%          \\
Tracers in  &  1000 & 0.1 $< t <$ 2 s & 0.39 & 0.50 & 0.44 & 0.31 $\pm$ 0.01 & 0.33 $\pm$ 0.02        & 6\%           \\
\;\;\;untreated cells       &       & 2 $ <t <$ 8  s  & 0.60 & 0.50 & 0.47 & 0.55 $\pm$ 0.01 & 0.57 $\pm$ 0.01        & 4\%          \\
% telomeres      & 0.12 $< t <$ 10 s & 0.32 & 0.50 & 0.52 & -- & 0.34 $\pm$ 0.01        & --            \\
%               & 40 $< t < 200$  s& 0.47 & 0.50 & 0.47 & -- & 0.44 $\pm$ 0.01        & --               \\
Amoeba      &  1142 & 1 $< t <$ 6 min & 0.61 & 0.50 & 0.44 & 0.58 $\pm$ 0.01 & 0.55 $\pm$ 0.03        & 5\%       \\
            &   & 10 $< t <$ 100 min & 0.52 & 0.52 & 0.37 & 0.42 $\pm$ 0.02 & 0.40 $\pm$ 0.02        & 5\%      \\
Harvester ants  &    67   & 10 $< t <$ 100 s & 0.88 & 0.50 & 0.57 & 0.92 $\pm$ 0.12 & 0.95 $\pm$ 0.08        & 3\%       \\
    &          & 100 $< t <$ 400 s & 0.59 & 0.51 & 0.35 & 0.47 $\pm$ 0.03 & 0.45 $\pm$ 0.03        & 4\%      \\
% mRNAp          & short times & 0.75 & 0.50 & 0.51 & 0.74 $\pm$ 0.11 & 0.76 $\pm$ 0.02        & 0.03    &        \\
%               & long times  & 0.52 & 0.49 & 0.53 & 0.53 $\pm$ 0.02 & 0.54 $\pm$ 0.01        & 0.02   &        \\
Commuting kite & 107 & $0.1  < t < 3$ min & 0.87 & 0.50 & 0.49 & 0.84 $\pm$ 0.02 & 0.86 $\pm$ 0.02        & 2\%         \\
        &         & $3 < t <  12$ min  & 0.80 & 0.50 & 0.50 & 0.76 $\pm$ 0.01 & 0.80 $\pm$ 0.02        & 5\%          \\
Searching kite & 587 & $0.5< t < 20$ min  & 0.24 & 0.59 & 0.22 & 0.06 $\pm$ 0.02 & 0.06 $\pm$ 0.01        &    $<$1\%  \\  
Stork (Jun-Jul) &  687 & $0.2 < t < 2$ h  & 0.43 & 0.85 & -0.22 &  0.07  $\pm$ 0.03   &   0.06 $\pm$ 0.04    &   **    \\
 &  & $2 < t < 10$ h & 0.13 & 0.55 & 0.42 &  0.13 $\pm$ 0.06    & 0.10 $\pm$ 0.01     & **      &    \\
Stork (Aug-Sep) &  165 & $0.2 < t < 4$ h   & 0.97 & 0.50 & 0.71 & 1.18 $\pm$ 0.01 & 1.18 $\pm$ 0.06      & $<$1\%        \\
%  &   & $2 < t < 10$ h & 0.84 & 0.50 & 0.70 & 1.03 $\pm$ 0.01 & 1.04 $\pm$ 0.06      & 1\%       \\
Stork (Oct-Jan) & 810 & $0.2 < t < 4$ h   & 0.70 & 0.62 & 0.27 & 0.56 $\pm$ 0.01 & 0.59 $\pm$ 0.03      & 5\%       \\
% &  & $2 < t < 10$ h   & 0.57 & 0.50 & 0.65 & 0.80 $\pm$ 0.02 & 0.73 $\pm$ 0.10      & 9\%         \\
Stork (Mar-Apr) & 255 & $0.2 < t < 4$ h   & 0.97 & 0.50 & 0.70 & 1.23 $\pm$ 0.01 & 1.18 $\pm$ 0.06      & 4\%          \\
%  &  & $2 < t < 10$ h   & 0.80 & 0.50 & 0.65 & 1.01 $\pm$ 0.01 & 0.95 $\pm$ 0.07      & 6\%          \\
 Vulture     & 444  & $0.1 < t < 2$ h   & 0.75 & 0.50 & 0.58 & 0.86 $\pm$ 0.02 & 0.84 $\pm$ 0.02      & 2\%     \\
 &      & $2< t< 5.5 $ h   & 0.56 & 0.50 & 0.63 & 0.64 $\pm$ 0.02 & 0.69 $\pm$ 0.04      & 8\%          \\
               [0.5ex]
\hline
%&&&\multicolumn{3}{c}{\textbf{simulations}}  &&&&     
\textbf{Simulations} & ensemble size &  parameters &{}  & {} & {} & {}& {}      & {}      &   {}  \\
%[-2.5ex]
\hline
Brownian motion     &    &  & 0.50 & 0.50 & 0.50 & 0.50 & 0.50       &    $<$1\%       \\
SBM     &  $10^5$  &  M = 0.3 & 0.50 & 0.50 & 0.30 & 0.30 & 0.30       &      $<$1\%       \\
LF       & $10^5$ &  L = 0.71 & 0.50 & 0.71 & 0.50 & 0.71 & 0.71   &        $<$1\%    \\
FBM      & $10^5$  &  J = 0.3 & 0.30 & 0.50 & 0.50 & 0.30 & 0.30       &     $<$1\%      \\
FBM     &  $10^5$  &  J = 0.7 & 0.70 & 0.50 & 0.50 & 0.70 & 0.70        &       $<$1\%    \\
SFBM    &   $10^5$  &  J=M=0.7 & 0.70 & 0.50 & 0.70 & 0.90 & 0.90        &     $<$1\%        \\
SFLM    &   $10^5$  &  J=L=0.6, M=0.3 & 0.60 & 0.60 & 0.30 & 0.50 & 0.50     &   $<$1\%   \\
% CTRW   &      & $\alpha = 0.7$ & 0.49 & 0.67 & 0.15 & 0.29 $\pm$ 0.01 & 0.31 $\pm$ 0.02        &        \\
CTRW     & $10^3$  & $\alpha = 0.8$ & 0.50 & 0.62 & 0.27 & 0.38 $\pm$ 0.01 & 0.38 $\pm$ 0.01        &      $<$1\%     \\
CTRW     &  $10^3$  & $\alpha = 0.4$ & 0.50 & 0.80 & -0.11 & 0.21 $\pm$ 0.01 & 0.20 $\pm$ 0.01        &   5\%    \\
BoCTRW   &  $10^3$ & $\alpha = 0.7$ & 0.15 & 0.66 & 0.18 & 0 $\pm$ 0.01 & -0.01 $\pm$ 0.01        &    **   \\
% BoCTRW & $\alpha = 0.5$ & 0.22 & 0.75 & 0.01 & 0 $\pm$ 0.01 & -0.02 $\pm$ 0.01        &  *  \\
BiCTRW & $10^3$ & $\alpha = 0.8$ & 0.92 & 0.63 & 0.24 & 0.75  & 0.78        &   4\%   \\
BiBM & $10^3$ &  & 1 & 0.5 & 0.5 & 1  & 1        &   $<$1\%     \\
[0.5ex]
\hline\hline 
\end{tabular}
\caption{\footnotesize{Summary of the decomposition of the origins of anomalous diffusion, in various data sets and simulations. We present the evaluated error on both $H$ and $H_p$. We also present the relative error between the directly-measured value of $H$ and its prediction via the sum rule \eqref{summation_relation}, $|H_p-H|/H$ in percentages.  In cases where the relative error of either the observed or predicted values exceeds the difference between the two (marked by the ** symbol), we assume good agreement regardless of the relative error, which can naturally be large for small values of the Hurst exponent. 
The simulations represent some prototypical examples of anomalous processes, and obey the summation relation. CTRW simulations were done in the current study, see Appendix~\ref{SubSecCTRW} and Sec. S3 in~\cite{SM}, and all other simulated systems are results quoted from Ref.~\cite{chen2017anomalous}. Legend: BM = Brownian Motion. SBM = Scaled BM. LM = L\'evy Motion (a.k.a., L\'evy flight). FBM = Fractional BM. SFBM = Scaled FBM.  SFLM = Scaled Fractional LM. CTRW = Continuous-Time Random Walk. BoCTRW = Bounded CTRW. BiCTRW = Biased CTRW. BiBM = Biased BM. 
Note that the errors in simulations quoted from~\cite{chen2017anomalous} are less than $10^{-4}$. }}
\label{tab:exponents} 

\end{tcolorbox}
 \end{table*} 
 \thispagestyle{empty}
%%%%%%%%% End Table 

\subsection{ Rhodamine molecules}
For the fluorescent rhodamine molecules~\cite{sarfati2020temporally}, we have detected a Joseph effect (anti-correlation), leading to subdiffusion (Sec.~S1.1 in~\cite{SM}), where the effect is strongest at the lowest relative humidity of 30\%  ($J=0.09$), see Table~\ref{tab:exponents}. 
This effect can be interpreted as viscoelasticity in the water film, possibly due to significant persistent ordering between H$_2$O molecules induced by strong coupling to the silicate or silanol groups of the silica surface. As humidity increases and water nanofilms grow in thickness, molecular ordering becomes more random and less persistent, and the water film becomes more viscous farther from the silica surface, resulting in a decrease of $J$.
A fractional Brownian motion (FBM)-like process~\cite{metzler2014anomalous} can be used to model the movement of these particles, exhibiting confined diffusion. Yet, contrary to ``pure" FBM, we  also observed a negative (although weak) Moses effect at short times for any humidity, suggesting a combined effect of FBM with SBM or continuous-time random walk (CTRW)~\cite{metzler2014anomalous}, see simulation results in Table~\ref{tab:exponents}. 
The subordination of FBM by a CTRW is also suggested by~\cite{sarfati2020temporally} and is consistent with the physical mechanisms of intermittent diffusion at solid/liquid interfaces, whereby a molecule desorbs from the surface, diffuses in the viscous phase, and re-adsorbs~\cite{skaug2013intermittent, wang2020non}. 

\subsection{ Tracer particles in mammalian cells}
%--> text fragment for Results, page 5
%Single-particle tracking data were obtained for quantum dots that had been introduced into the cytoplasm of treated and untreated mammalian cells~\cite{sabri2020elucidating}. 
%In particular, we analyzed large ensembles of tracks obtained with a frame time $\Delta = 100$~ms in untreated cells and in cells that have been treated with latrunculin~A to disrupt the actin cytoskeleton. The selected tracks in untreated cells had a length of $N=100$ positions whereas the chosen trajectories for latrunculin-treated cells each had $N=500$ positions. 
In both treated and untreated cells~\cite{sabri2020elucidating}, the statistics display two temporal regimes (Sec.~S1.2 in~\cite{SM}). In the first regime ($t<5$ s and $t<2$ s for the treated and untreated cells, respectively) the dynamics are anti-correlated, with a weak negative Moses effect and no Noah effect. Together, these effects lead to significant subdiffusion with $H = 0.31$. 
In contrast, in the second regime ($t>5$ s and  $t>2$ s) the Joseph effect is measurably different between treated and untreated cells. While for untreated cells the dynamics are positively correlated and hence superdiffusive, for the treated cells they are not correlated, and a Moses effect leads to subdiffusion. Our results are consistent with those of~\cite{sabri2020elucidating} (Sec.~S1.2 in~\cite{SM}).
The elevated values for $J$ in the second regime suggest that on time scales of a few seconds particles are being kicked by an active ambient noise that arises by cytoskeleton-associated transport processes in the surrounding~\cite{guo2014,stadler2017non, SSW2018}. 
%,}. 
This notion is in line with a reduction of $J$ when breaking down actin filaments which prevents the contribution of slow active processes linked to cell reshaping and migration. The presence of a weak Moses effect at all time scales is most likely due to the intermittent mobility change found for these tracers, as they transiently adsorb to and desorb from the cell's vast endomembrane system~\cite{sabri2020elucidating}.
Notably, in temporal regimes with measurable Joseph and Moses effects, the system can be modeled, \textit{e.g.}, by scaled FBM (SFBM), see Table~\ref{tab:exponents}.
Yet, as the Moses effect is very weak, and as particles are non-specifically bound to a dynamic endoplasmic reticulum~\cite{sabri2020elucidating}, FBM cannot be discarded entirely.

%%%%%%%%%%%%%%%%%%%%%%%%%%%%%%%%%%%%%%% Figure 3: 
\begin{figure*}[t]
\centering
\includegraphics[width=1\linewidth]{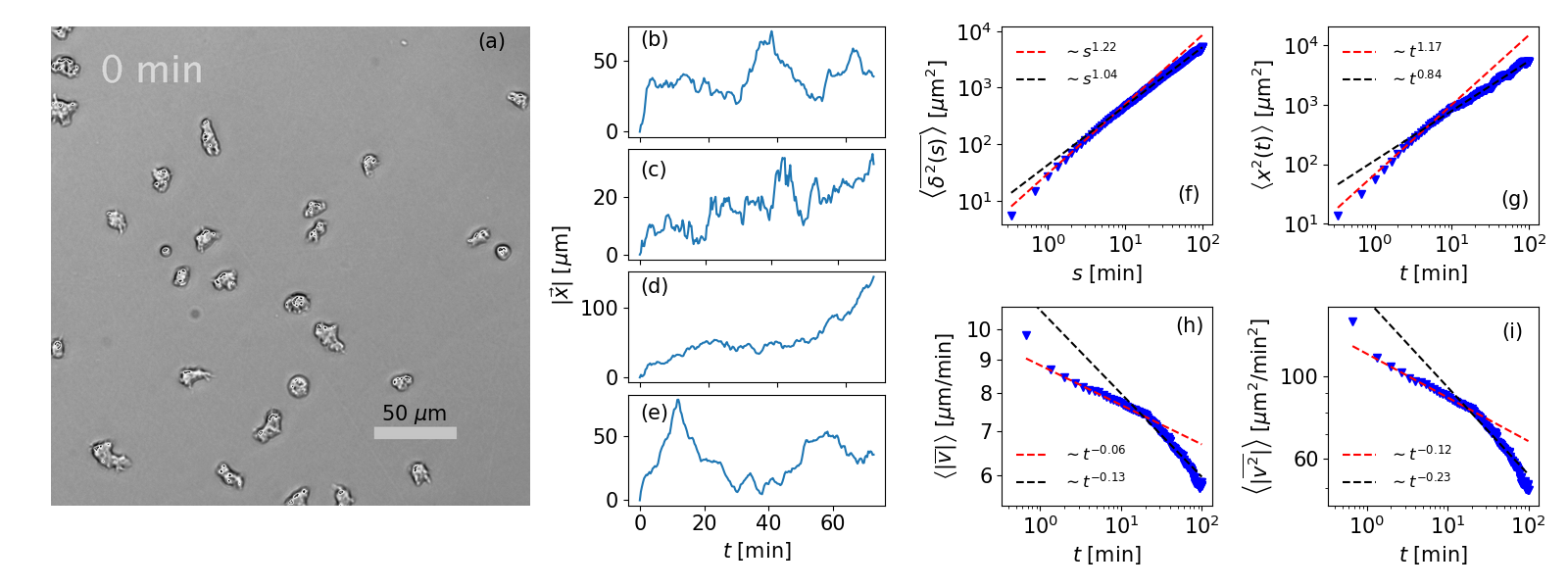}
 \caption {\footnotesize{Amoeba: from raw data to statistics -- depiction of the workflow repeated for all empirical setups. Panel (a) is a snapshot of tracked amoeba, measured experimentally in~\cite{cherstvy2018non}. From the 1142 two-dimensional paths we generate vector trajectories for $\mathbf{x}(t)$ [setting $\mathbf{x}(0)=0$],  which we decompose into increments (see Fig.~\ref{fig:Scheme}) and obtain one-dimensional paths indicating the distance traveled versus time [panels (b-e)]. For each time series and measurement time $t$ we compute the TAs and EAs of the squared-displacement $\delta^2(s)$ (f), absolute-velocity  $|\mathbf{v}(t)|$ (h), and squared-velocity $\mathbf{v}^2(t)$ (i). To independently measure $H$, we further obtain the MSD $\langle\mathbf{x}^2(t)\rangle$ (g). Using the increment statistics in panels (f-i) we visually identify two regimes, $1 < t< 6$ min and $10 < t < 100$ min, that fit local scaling exponents (within a certain finite period of time, see Appendix~\ref{SubSecLocalScaling}). Each regime is then fitted with a power-law using the method of nonlinear least-squares. The measured values are marked in blue triangles and the fits are plotted as red and black dashed lines for the first and second regimes, respectively. The fit values are given in the legends, and are used to extract the scaling exponents $J, L, M$ and $H$. Specifically, $J$ and $H$ are extracted from panels (f) and (g) respectively,  $M$ is extracted from panel (h), and $L$ is extracted from panel (i) using the previously found value of $M$ [see Eq.~(\ref{eq2})].}}
 \label{fig:amoeba}
\end{figure*}

\subsection{ Amoebas}
In Fig.~\ref{fig:amoeba} we plot the results for the tracked amoebas~\cite{cherstvy2018non}, and depict the analysis workflow that we repeat for all empirical data sets.  
Here, the statistics display two temporal regimes, with different scaling exponents, which indicates a change in the amoeba dynamics at intermediate times.
For $1<t<6$ min the dynamics are positively correlated ($J = 0.61$), which is the dominant effect leading to superdiffusion ($H>0.5$), and exhibit a weak Moses effect and no Noah effect. In contrast, for $10<t<100$ min the dynamics are not correlated; rather, a negative Moses effect, entailing statistical slowing down, leads to subdiffusion ($H<0.5$). Thus, the primary cause for anomalous diffusion differs between the regimes. In the first regime we conjecture that the underlying process is FBM with positive correlations, while in the second regime it resembles SBM resulting in a Moses effect, in agreement with the analysis performed in Ref.~\cite{cherstvy2018non}. A consistency check of the observed exponents with concrete stochastic simulations is given in Table~\ref{tab:exponents}, while in Appendix \ref{appendix:p} and Fig. S14 in~\cite{SM}, we provide independent validation of the above results using a p-variation test~\cite{magdziarz2009fractional, meroz2015toolbox}.  Importantly, while the extracted exponents do not allow unique model identification, they provide crucial insights into the detailed dynamics of the observed motion.

\subsection{Ants}
For the harvester ants~\cite{avgar2008linking}, we find (Sec.~S1.3 in~\cite{SM}) that for $10<t<100$ s the movement is strongly correlated ($J = 0.88$) with a small positive Moses effect, leading to superdiffusion. In this regime the ants are  behaviorally persistent, primarily commuting between the nest and food sources in relatively straight lines, leading to biased-correlated movement. In contrast, for $100<t<400$ s the movement is less correlated and non-stationary ($J = 0.59$ and $M = 0.35$). Both the positive Joseph and negative Moses effects likely reflect behavioral shifts between commuting (superdiffusive) and searching or handling seeds (diffusive or subdiffusive). While the Hurst exponent may suggest (almost) Brownian diffusion at these times, this is not the case; rather we measure $H = 0.47$ due to a nontrivial coupling of the Joseph and Moses effects, likely common in many central-place foraging movements. 

\subsection{\ Kite}
\vspace{-3mm}We separately analyse ensembles of commuting and search flights~\cite{vilk2021ergodicity}. During commuting, for $t<3$ min and $t>3$ min the dynamics are positively correlated ($J = 0.87$ and $J = 0.80$ respectively), leading to superdiffusion (Sec.~S1.4 in~\cite{SM}). Here, the Moses and Noah effects are negligible, and since the MSD and  TAMSD scale similarly with time, the process is ergodic~\cite{metzler2014anomalous}.  Indeed, the most efficient way to commute among patches is to fly in a straight line directed towards the target (strong positive Joseph effect). These commuting flights occur at a steady cruising speed (no Moses effect) and also without extreme jumps (no Noah effect), suggesting lack of support for the L\'evy foraging hypothesis~\cite{benhamou2007many, benhamou2014scales, spiegel2015moving}. %The long-range correlations are expected for a commuting animal, attempting to cover large distances during flights. Yet, for long times the correlations are weaker. This may occur as some long commuting flights do not reflect ballistic motion towards a target, but rather long-range exploration which is less directed at long times. 
During searches, the statistics display a single regime (Sec.~S1.4 in~\cite{SM}). Here the dynamics are anti-correlated with $J = 0.24$ and there are measurable Moses and Noah effects, $M = 0.22$ and $L = 0.59$. Kites search locally in a spatially confined manner to avoid departure from a patch (negative Joseph
effect), with relatively long stops in particular locations (negative Moses effect) and also rare long jumps between these locations (Noah effect). Our results support Ref.~\cite{vilk2021ergodicity} that the kite's searches can be modeled as a bounded CTRW, see Table~\ref{tab:exponents}~\footnote{For additional tests showing that the searching kite can be modeled as a bounded CTRW, see Ref.~\cite{vilk2021ergodicity}. In Appendix~\ref{appendix:p} and Fig. S15 in~\cite{SM} we further provide  a p-variation test for randomly chosen trajectories.}.

\begin{figure*}[t!]
\centering
\includegraphics[width=0.95\linewidth]{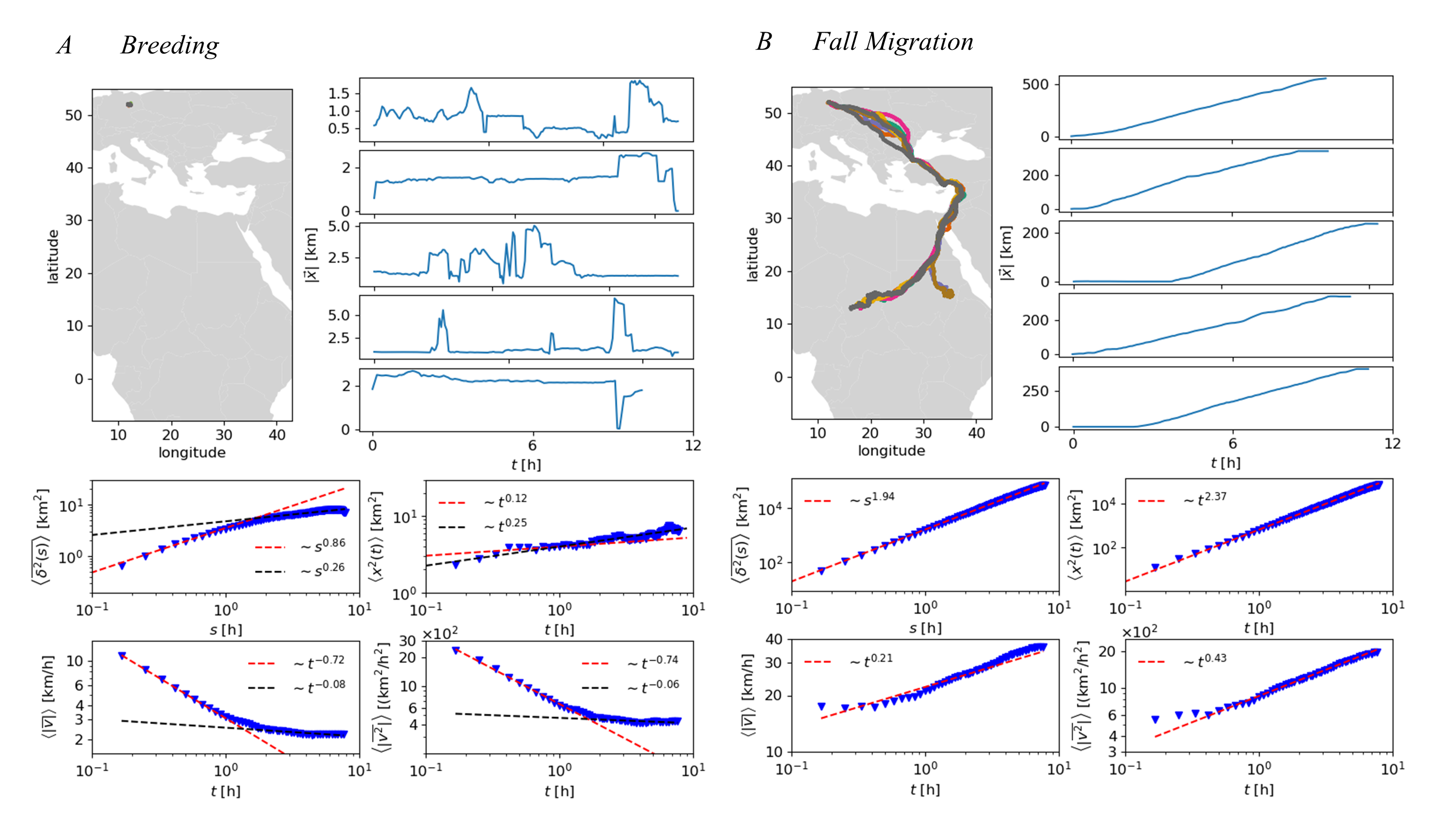}
 \vspace{-9mm}
 \caption {\footnotesize{Stork: from raw data to statistics during June-July (A, breeding, 687 trajectories) and August-September (B, fall migration, 165 trajectories). In both A and B, the upper left panels are GPS tracks, where different colors represent different years. The five upper right panels are examples for the distance traveled at a single day versus the measurement time in hours. The four lower panels are the statistics on the ensemble of days during the relevant period, and the fit values are given in the plot legends (see Fig.~\ref{fig:amoeba} for details). 
 }}
 \label{fig:storkGPSplot}
\end{figure*} 

\subsection{Stork}
\vspace{-3mm}Daily paths are clustered into four subsets, based on the time of year: June-July, August-September, October-January, March-April, respectively corresponding to four periods in a bird's life cycle: breeding, fall migration, wintering, and spring migration~\cite{rotics2016challenges}. During each of the above periods we analyze the subset of days with total displacement that is consistent with the assumed behavior (e.g., for a migrating bird we only analyze paths with total displacement $> 100$ km). With this simple clustering we aim to capture important features of the stork's life history (Fig.~\ref{fig:storkGPSplot} and Fig. S11 in~\cite{SM}). %Notably, February and May were omitted from our analysis as during these months the paths do not strictly correspond to either breeding, wintering or migrating period. 

During \textit{breeding} (June-July) we observe subdiffusive motion  at all times ($H<0.5$). For $t<2$ h and $t>2$ h  subdiffusion is caused by a strong negative Moses effect coupled to a strong Noah effect ($M = -0.22, L = 0.85$), and anti-correlated movement ($J = 0.13$), respectively. In hot  days, breeding storks fly early in the morning to forage in neighboring fields ($>5$ km away~\cite{rotics2016challenges}), but remain longer times in the nest during the hottest hours to thermoregulate the eggs or nestlings. Flights occur earlier in the morning, and are much faster and less frequent than foraging walks or stops in the nest, with relatively long waiting times, explaining the strong Noah and negative Moses effects in the first regime. The tendency to return to the nest during the remaining parts of the day explains the strong negative Joseph effect appearing in the second regime, while the disappearance of the Moses and Noah effects in this regime may be since the waiting time distribution is no longer scale free.
Notably, the coupling of the Moses and Noah effects was also found, \textit{e.g.}, in CTRW simulations,  see Table~\ref{tab:exponents} and Appendix~\ref{SubSecCTRW}, and is consistent with known theoretical results for CTRW ~\cite{aghion2021moses,Meyer2021Decomposing}, see below. In contrast, the observed exponents in the second regime may emanate from movement within a (self determined) bounded domain, or from FBM, see Table~\ref{tab:exponents}. 
Both long waiting times and bounded movement are supported by the movement paths (Fig.~\ref{fig:storkGPSplot}A and Appendix~\ref{appendix:p}) and may be generated by the spatio-temporal constraints of a breeding animal.

\begin{figure*}[t!]
\centering
\includegraphics[width=0.95\linewidth]{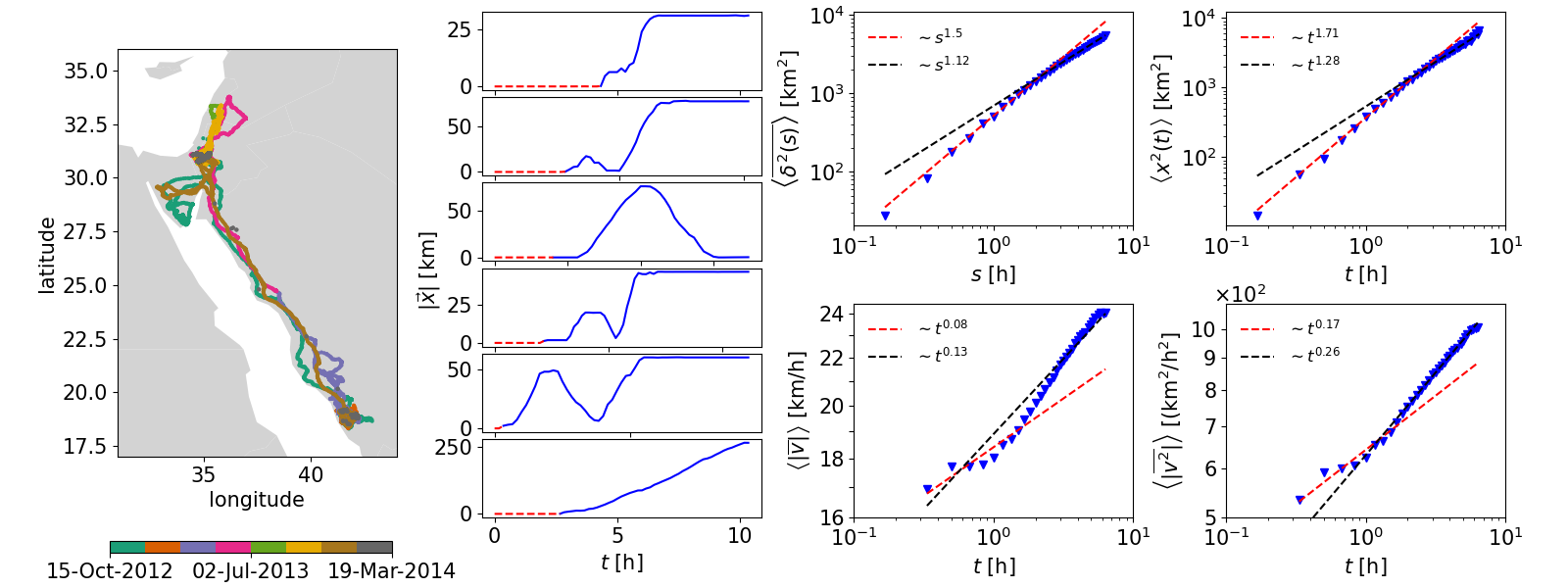}
 \vspace{-2mm}
 \caption {\footnotesize{Vulture (444 trajectories): from raw data to statistics. Left panel: GPS tracks of the vulture. Middle panels: examples for the distance traveled (km) versus time (hours).  As was done in~\cite{harel2016decision}, the individual paths are normalized to start at the first relocation in the day with a velocity $>$4 m/s, as to avoid aging of the system due to long initial waiting times. The red dashed lines represent initial stationary periods of the animal. Four right panels: statistics fitted in two regimes,  $0.1<t<2$ h and  $2<t<5.5$ h (see Fig.~\ref{fig:amoeba} for details).
 } }
 \label{fig:vultureGPSplot}
\end{figure*} 

During \textit{wintering} (October-January) the movement patterns are superdiffusive ($H = 0.56$). Here, superdiffusivity is primarily caused by long-range correlations ($J = 0.70$) and a Noah effect ($L = 0.62$), which are balanced with a negative Moses effect ($M = 0.27$), such that the movement is evidently nonergodic. Note that for times $t > 4$ h the statistics do not display a clear power law (see Fig. S11A in~\cite{SM}).
These movement patterns reflect a mixture between the stork's moves during breeding and migration, see below. On the one hand, during winter they move much longer distances than during breeding, including long migration-like directional flights to distant wintering sites (Fig.~S11A in~\cite{SM}) that give rise to a positive Joseph effect. On the other hand, wintering storks resemble breeding ones in their confined movement for days or weeks and roosting in a central place from which they fly to their foraging sites early in the morning. In these sites they search for food mostly by walking. The combination of long-range flights and local walks gives rise to Noah and negative Moses effects (as in breeding, see above). %Storks also tend to be less mobile during mid day (as in breeding) but resume foraging in the afternoon (unlike breeding), explaining the lack of Noah effect and the shift to a positive Moses effect in the second regime.
% The fact that during winter the stork displays superdiffusion may reflect the lack of an evolutionary constraint to remain at a specific location during winter, coupled to a possible nomadic tendency during long periods in the winter. Indeed, direct observations of the GPS tracks in Fig.~\ref{fig:storkGPSplot} reveal that in contrast to breeding, while wintering the stork covers much more ground and may commute between remote regions.  
The Noah effect ($L>1/2$), appearing during both breeding and wintering, may stem from the fact that various aspects of the stork's daily routine remain constant, despite the underlying seasonal behavior. Throughout the year the stork can move via walking at a range of velocities~\cite{van2004terrestrial},
short-term flights (mostly within a food patch) and long-term commuting (\textit{e.g.}, from the nest to foraging ground). Thus, flights may appear as a heavy tail compared to the bulk, comprised mainly of short-scale walks. This is in striking contrast to the lack of a Noah effect for the vulture which travels only via flights, see below.

During fall and spring \textit{migrations} (August-September and March-April) the statistics are similar: for $t\!<\!2$ h and $t>2$ h we find  strongly correlated movement with a small positive Noah effect, and a strong  positive Moses effect with no Noah effect, respectively.
Migrating storks take highly directional flights from the breeding to wintering grounds during fall (Fig.~\ref{fig:storkGPSplot}B), and vice versa during spring (Fig.~S11B in~\cite{SM}), giving rise to a strong positive Joseph effect in both cases. They roost in stopover sites during night and tend to depart in late morning, when soaring conditions improve, facilitating faster flights at lower energy costs~\cite{horvitz2014gliding}. This explains the positive Moses effect in the second regime. %The long correlations are a result of long distances covered during migration (see Fig.~\ref{fig:storkGPSplot}), and are clearly interpreted as biased movement towards breeding or wintering grounds. Interestingly, at long times we find a no Noah effect and strong non-stationarity in the form of a positive Moses effect. This may reflect acceleration of the migrating animal that may have rested during the day, to cover long distances at later times; however, further modeling efforts are required to test such a hypothesis. 
Here, a plausible model for movement is a scaled FBM~\cite{bel2005weak}, see Table~\ref{tab:exponents} and Appendix \ref{appendix:p}.

%Importantly, for all these subsets, the Noah and Moses exponents display similar behavior: $M < 0.5$ and $L > 0.5$ in the first regime and $M > 0.5$ and $L \simeq 0.5$ in the second regime. We suggest that while the animal is driven by seasonal behavior, various aspects of its daily routine remain constant. For example, throughout the year the stork can transport via walking at a range of velocities~\cite{van2004terrestrial}, short-term flights (mostly within a food patch) and long-term commuting (e.g., from the nest to foraging ground). Such a range of behaviors across multiple scales can give rise to a measurable Noah effect, as flights appear as a heavy tail compared to the bulk, comprised mainly of short-scale walks. This is in striking contrast to the vulture, see below, which travels only via flights and accordingly does not display a measurable Noah effect.

%%%%%%%%%%%%%%%%%%%%%%%%%%%%%%%%%%%%%%% Figure 3: 

\subsection{ Vulture}
For the daily paths of the vulture~\cite{harel2016decision}, %the statistics are averaged over an ensemble of 429 two-dimensional daily paths, each measured at a frequency of 1/600 Hz for up to 11 hours~\cite{harel2016decision}, see Fig.~\ref{fig:vultureGPSplot}. 
for $t<2$ h the movement is superdiffusive and ergodic ~\cite{metzler2014anomalous, mangalam2021point}, as it is positively correlated ($J = 0.75$), with a weak Moses and no Noah effects; for $t>2$ h a positive Moses effect ($J = 0.56$ and $M = 0.63$) leads to superdiffusive behavior and ergodicity breaking. %In both regimes the Noah effect is negligible. 
Vultures fly relatively straight away from, or back to, their roost, and to search for occasional carcasses or those randomly (in time) supplied in a few dozens of feeding stations scattered throughout their foraging area in Israel, explaining the positive Joseph effect.  Despite the occurrence of very long flights (Fig.~\ref{fig:vultureGPSplot}), the L\'evy foraging hypothesis is not supported for this species (no Noah effect), in accord with~\cite{spiegel2015moving}. Vultures tend to move faster towards a known target 
%(e.g. detected carcass, back to the roost) 
compared to the preceding search phase~\cite{harel2016decision}, and like migrating storks, they fly faster when soaring conditions improve (from late morning to early afternoon), altogether explaining the positive Moses effect.    
%entailing that the observed movement does not follow L\'evy flight dynamics. Rather, at short times, the superdiffusive behavior emanates from long-range temporal correlations, which could be interpreted in one of two ways: (a) biased movement of the vulture, flying towards a designated location (e.g., nest or feeding station), or, (b) correlated movement for a randomly foraging vulture~\cite{johnson2008continuous}.  These results have significant implications on the modeling of these tracks, suggesting that a combination of biased-correlated random walks and SBM, see Table.~\ref{tab:exponents}, are better suited  to model this movement process than L\'evy flights~\cite{spiegel2015moving}. 

%%%%%%%%%%%%%%%%%%%%%%%%%%%%%%%%%%%%%%% Figure 5:

\begin{figure*}[t!]
\centering
\includegraphics[width=0.82\linewidth]{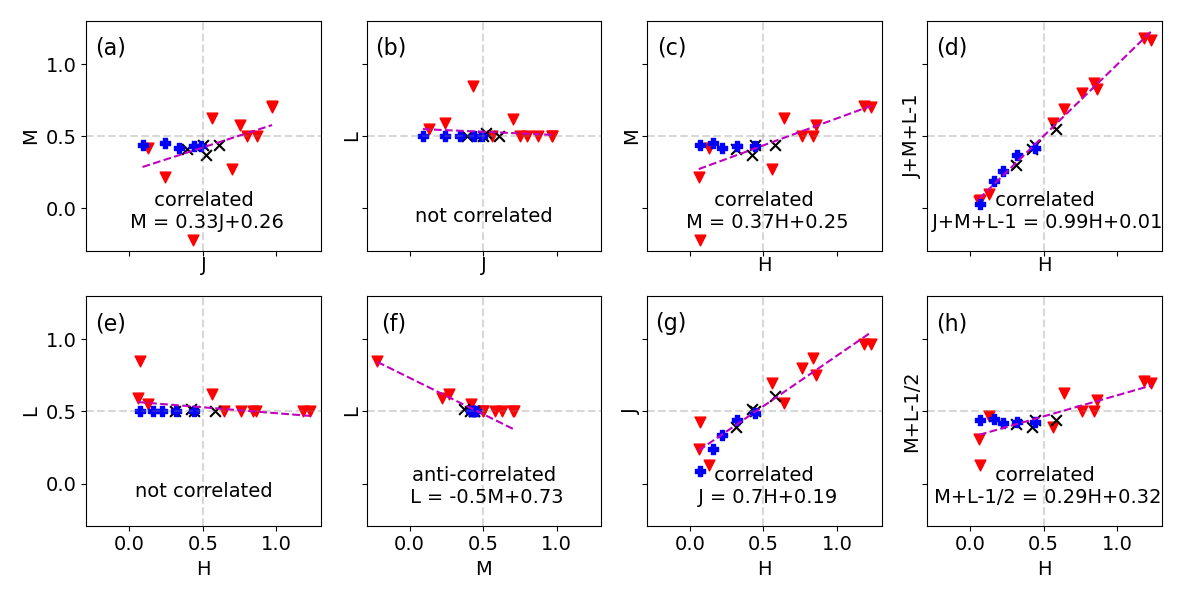}
\vspace{-7mm}
 \caption {\footnotesize{Correlations between any two exponents measured for the  empirical systems (values taken from Table \ref{tab:exponents}), and the relation between $H$ and the prediction of the summation relation. Blue pluses mark the chemical data sets, black crosses mark the biological data sets, and red triangles mark the ecological data sets. The size of each marker reflects a typical (average) error of 0.03. The purple dashed line represents a linear least square fit. The data was assumed to be correlated (anti-correlated) if Pearson's correlation coefficient test showed a p-value $< 0.05$ and yielded a positive (negative) test statistics. For (anti-)correlated fits the results of the linear fit are explicitly written. Note that for panel (f) the fit was performed only for the non-trivial points for which $L > 0.5$ (see text). %\red{[[Maybe also say $M<1/2$? Is this correct (see remark from the referee)]]}. %Note, that we have omitted two points from the analysis (a stork in the breeding season), since the $H$ value there could not be determined.
 }}
 \label{fig:exponentsMap}
\end{figure*}

%%%%%%%%%%%%%%%%%%%%%%%%%%%%%%%%%%%%%%%
%  Section: Discussion 
%%%%%%%%%%%%%%%%%%%%%%%%%%%%%%%%%%%%%%%
\section{Summary and discussion}
\label{SecDiscussion} 
We have demonstrated the wide applicability of a general method~\cite{mandelbrot1968noah,chen2017anomalous} to unravel the origins of anomalous transport in empirical time series in chemistry, biology and ecology, over multiple spatio-temporal scales.
%and provided a framework to analyze data from a large variety of sources. 
Using positional time series, almost free from prior assumptions and with little-to-none auxiliary information, the method decomposes the Hurst exponent into three components: nonstationarity, fat-tailed distributions, and long-range correlations. The decomposition is manifested by a summation rule [Eq.~\eqref{summation_relation}], and is verified for all analyzed data sets. We stress that although the summation rule was conjectured and applied in a number of previous studies, our study is the first to empirically test this conjecture over a wide range of data, thus confirming its validity.

Previous works have shown that in several models the exponents can be interconnected~\cite{chen2017anomalous,meyer2018anomalous,aghion2021moses}, and processes can be associated with multiple effects. Indeed, our analysis points to inherent correlations and physical differences between the analysed scaling exponents. In Fig.~\ref{fig:exponentsMap} we study these correlations by plotting the relation between various combinations of the exponents $J, L, M$ and $H$, using the values in Table~\ref{tab:exponents}. These relations allow us to conjecture regarding the interplay between the exponents in real-life processes. 
In particular, we find that $J$ and $H$ (Fig. \ref{fig:exponentsMap}g) are strongly  correlated, with $J \sim 0.70 H$ when measured from the ensemble of all datasets; yet, $J$ alone is not sufficient to predict the value of $H$. Rather, we find that $J + M + L - 1 \simeq 0.99 H$, in excellent agreement with Eq. \eqref{summation_relation}. The fact that $J$ is generally smaller than $H$ indicates that,  while some anomalous processes are ergodic, many are not. Thus, the ergodicity assumption may lead to erroneous analyses~\cite{mangalam2021point}. 
Figure~\ref{fig:exponentsMap} also reveals that $M$ and $L$ are anticorrelated (Fig.~\ref{fig:exponentsMap}f, $L \sim -0.5 M$), while $M$ and $J$ are correlated (Fig.~\ref{fig:exponentsMap}a, $M \sim 0.33J$).  This suggests inherent relations between the Moses and Noah (especially evident in the ecological and biological data), and Moses and Joseph effects.  
This entails that among the systems we study, nonstationarity (e.g., $M < 0.5$) is not likely to be pure SBM; rather, the process will also exhibit correlations ($J < 0.5$) and/or a fat-tailed distribution ($L >0.5$). 
The link found between $M$ and $L$, which primarily emanates from the ecological data sets, reproduces the known relation found in CTRW~\cite{aghion2021moses,Meyer2021Decomposing}: $L=-M/2+3/4$ (compare to  $L\!=\!-M/2\!+\!0.73$ in Fig.~\ref{fig:exponentsMap}f), and supports the suggestion that the processes with $L>1/2$  may be described by (anti)correlated CTRW. Finally,  considering the whole ensemble of data sets, we find that $L$ and $H$ are uncorrelated, suggesting that L\'evy-flight-like processes are rarer, as $L$ is subdominant compared to the other exponents.

Although  no unique model can be assigned to a system based only on its given set of exponents, our framework plays a key role as a decision tree allowing to identify a model class, and rule out inappropriate models~\cite{meroz2015toolbox}. For example, in L\'evy flights as defined in Ref.~\cite{chen2017anomalous}, one expects $L>1/2$ and $M=J=0.5$, which was not found for any of the data sets we analyzed. Instead, as shown above, CTRW is a more plausible model for searching kites and breeding storks. This finding gives key insight into an open question in ecology, of whether, for a given data set, an animal follows a L\'evy flight
(Noah effect), or a combination of a biased-correlated random walk (Joseph effect) and scaled motion (Moses effect)~\cite{benhamou2014scales, auger2015differentiating}. 
%and thus, we can rule out the L\'evy flight model for vultures and  commuting kites, for which we have found $L = 1/2$. In these cases we rather believe that biased correlated random walks or fractional Brownian motion are more suitable models. Importantly, the  L\'evy flight model can also be ruled out  for the searching kite and stork during breeding or nesting, even though they do display a Noah effect with $L>1/2$. The reason is that this Noah effect is accompanied by a negative Moses effect ($M < 0.5$), which cannot occur for pure L\'evy flights~\cite{chen2017anomalous}, see Table~\ref{tab:exponents}. Yet, such behavior can be a result of a L\'evy \textit{walk} in certain parameter regimes~\cite{aghion2021moses} or a CTRW with long waiting times, see Table~\ref{tab:exponents}. 
Moreover, when the Joseph effect is present, specific empirical input is needed to distinguish between biased and correlated processes.
% For example, in Fig.~\ref{fig:vultureGPSplot} we see that on some days the vulture commutes outside Israel, hence the flights are clearly directed (biased). But, when the vulture remains confined within its home range, it is plausible that the animal is randomly foraging, and thus the Joseph effect represents correlated movement.
For example, within its large yet spatially confined foraging range, the vulture searches for carcasses in circular-like paths, whereas rare long-range forays outside its home range are highly directional~\cite{harel2016decision, spiegel2015moving} (Fig.~\ref{fig:vultureGPSplot}), suggesting that the Joseph effect represents, respectively, correlated and biased movement.

% Notably, while our framework can differentiate between L\'evy flight and biased-correlated random walks, we cannot determine whether a Joseph effect is due to biased motion to an a-priory known location or correlated movement, e.g. due to FBM~\cite{chen2017anomalous}. Here, further modeling efforts, or biological input, will play a crucial role.
%For instance, in Fig.~\ref{fig:vultureGPSplot} we see that in some days the vulture commutes outside Israel. In these cases the flights are clearly directed with a constant direction in sequential days, and can likely be described by a biased random walk. In contrast, for days in which the vulture remains within the home range, it is plausible that the animal is randomly foraging (and not moving towards a known location), and thus, it is plausible that the Joseph effect is due to correlated movement. 

%%% 
Three restrictions can strongly impact the data analysis in experiments involving anomalous diffusion.
First, as processes with a Moses effect are generally non-ergodic and can display aging~\cite{metzler2014anomalous}, the values of $M$ (and also $L$, see Fig.~\ref{fig:exponentsMap}) can change depending on the relative time lag between the process's initiation time and the initial measurement time. Thus, minimizing this time lag is desirable to reflect the properties of the measured phenomenon (Sec.~S2 and Fig. S12 in~\cite{SM}). %we show that artificially ageing the searching kite's ensemble, obscures important features of the system. 
Second, nonergodic processes are sensitive to the ensemble size, even for comparably large ensembles. As nonergodic systems display large variability across different realizations~\cite{metzler2014anomalous}, removing even a few can strongly affect the underlying statistics~\cite{mangalam2021point}. 
Third, results may be sensitive to the sampling frequency,
and in general, it is desirable to have a sampling frequency higher than the natural frequency of the process. 
In addition, when applying power-law fits to data, there are various statistical methods that provide confidence to the results. While $J, L$ and $M$ may be sensitive to the above restrictions and method of fitting,
we checked that they vary in such a way to maintain the validity of the summation relation. Regardless, in future theoretical work, it would be useful to generalize the theory [Eqs.~(\ref{eq0}-\ref{eq4})], to the case of non-pure power laws. %Importantly, as we have shown, applying this decomposition on empirical time series may provide means to resolve open questions in various scientific fields.

%\michael{\textbf{Move this paragraph somewhere else.}} {\color{red}Finally, when applying  power-law fits to data, one may use various methods and statistical tests to gain confidence about  the results, which are beyond the scope of the current work. In future theoretical work, it would be useful to re-examine the definitions of the exponents; Eqs. (\ref{eq1}-\ref{eq4}), in case of non-pure power-laws. } 

Notably, machine-learning algorithms, despite their ``black box" nature, may also be applicable to detect effects such as aging, extreme events and temporal autocorrelations. Indeed, in recent years there has been a growing effort in the scientific community to  advance the study of anomalous transport in data using a range of data-driven methods, such as machine-learning. %, see e.g.,~\cite{munoz2020single,ANDI2021,manzo2021extreme,gentili2021characterization,argun2021classification,Dezhong2021,pineda2022geometric}. 
Using massive training data sets, such algorithms often yield higher accuracy when extracting e.g., the Hurst exponent from single paths, or selecting between known stochastic models~\cite{ANDI2021}. 
In future works it would be very useful to generalize these algorithms and to allow for the estimation of the effects characterized by the three exponents $M$, $L$ and $J$, using data-driven algorithms. For example, aging effects can be  detected  via the power-spectrum of the time series~\cite{fox2021aging,vilk2022classification}, which can possibly be analyzed using machine-learning tools. Moreover, feature-based deep learning strategies \cite{kowalek2019classification} may profit from the three exponent decomposition, especially for $H\simeq 0.5$ \cite{loch2020impact}.

Finally, 
Based on the evidence and agreement of  our analysis, along 12 orders of magnitude in space and 8 orders of magnitude in time, we foresee that this method will provide useful results also in other fields such as cell biology and climate change, where anomalous time series are also common.

\vspace{-.3cm}
\section{Acknowledgment}
For fieldwork and technical assistance we thank Y. Serry (harvester ants), Y. Orchan and R. Shaish (kite), O. Spiegel and R. Harel (vulture) and S. Rotics and M. Kaatz (stork).
EA thanks Andrey Cherstvy and Kevin Bassler for useful discussions and  advice. CB acknowledges financial support from Deutsche Forschungsgemeinschaft (DFG), grant SFB1294/1-318763901.
A.S. and M.W. acknowledge support by the German Academic Exchange Service (PPP USA grant No. 57315749) and by the VolkswagenStiftung (Az. 92738). 
R.N. acknowledges support from  JNF/KKL grant 60-01-221-18, BSF grant 255/2008, and DIP (DFG) grant NA 846/1. RN also acknowledges support from Adelina and Massimo Della Pergola Chair of Life Sciences.  
R.M. acknowledges the German Science Foundation (DFG) for support within grant ME 1535/12-1.
O.V. and M.A. acknowledge support from the ISF grant 531/20. 
M.A. also acknowledges Alexander von Humboldt Foundation for an experienced researcher fellowship.

% \section{Author Contributions}
% \noindent OV, EA, RM and MA conceived and planned the study. OV, EA and MA performed the analytical calculations and analyzed the data. OV wrote the numerical code and performed the simulations. TA, CB, ON, AS, RS, DS, MW, DK and RN performed the experiments and collected the data. OV, EA and MA wrote the manuscript with valuable input from all other authors.

% \section{Competing Interests statement}
% \noindent  The authors declare no competing interests. 

\appendix
% \newpage %\onecolumngrid 
%This supplemental is organized as follows: Sec. \textbf{S1} contains 1. Details on the derivation of the Gaussian CLT, and the reasons why it is violated by the Moses,Noah and Joseph effects. 2. Details on the method of obtaining the Joseph exponent $J$. 3. A recipe for obtaining $M$ and $L$. Sec. \textbf{S2} contains technical details of the data analysis of experiments not covered in the main text.   

% \section{Statistical methods} \label{sec:MethodsSI}

%%%%% 
%%%%% Derivation of CLT  appendix - moved partially to the main text. 
\section{Derivation and violations of the Gaussian CLT}
\label{SubsecGaussianCLT}
We present here a well-known pedestrian derivation of the Gaussian  CLT~\cite{klafter2011first}, with emphasis on the assumptions of the theorem, which can be violated when the increments are non-stationary or long-ranged (or anti-) correlated, or if their mean-square is not finite. 

Consider, without loss of generality, the series of identically distributed random numbers $\delta x_0,\delta x_1,...\delta x_{n-1}$, with zero mean and variance $\sigma^2>0$ (which also equals the second moment in this case), as increments of the one-dimensional discrete process $x_n=x_0+\sum_{i=0}^{n-1} \delta x_i$. We define %$Y_n=\sum_{i=1}^{n-1}\delta x_i/\sqrt{n\sigma^2}$, and 
the probability density $W(\delta x_{i})\equiv W(x_{i+1}-x_{i})$, of traveling the  distance $x_{i+1}-x_{i}$, for $i=0..n-1$. If the increments are \textit{identically distributed}, and do not depend explicitly on location and time, the probability distribution $P(x,n)$ of being at $x_n=x$ after $n$ steps, is given by the recurrence equation $P(x,n)= \int_{-\infty}^\infty \mbox{d} x_{n-1} W(x_n-x_{n-1})P(x_{n-1},n-1)$.  %=\int_{-\infty}^\infty\mbox{d} x_{n-1}\int_{-\infty}^\infty \mbox{d} x_{n-2}  W(x-x_{n-1})W(x_{n-1}-x_{n-2})P(x_{n-2},n-2)$, and after many iterations; 
Initially we assume that $x_0=0$, namely $P(x,0)=\delta(x_0)$, where $\delta(\cdot)$ is the Dirac delta function. Since $x_0=x_n-\sum_{i=0}^{n-1} \delta x_i$, and $\delta(x)$ is symmetric, one can write  
\begin{eqnarray} 
P(x,n)&& = \int_{-\infty}^\infty\mbox{d} \delta x_{n-1}\int_{-\infty}^\infty \mbox{d} \delta x_{n-2}...\int_{-\infty}^\infty\mbox{d} \delta x_{0} \times \\&& W(\delta x_{n-1})W(\delta x_{n-2})...W(\delta x_0)\delta\left(x-\sum_{i=0}^{n-1} \delta x_i\right). \nonumber
\label{CLT1}
\end{eqnarray}  
%where the introduction of the Dirac Delta function $\delta(\cdot)$ accounts for the initial condition $P(x,0)=\delta[x_0]$. 
Defining the Fourier transform as  $f(x)\rightarrow\hat{f}(k)=\int_{-\infty}^\infty f(x)e^{ikx}\mbox{d} x$ and using the relation $\int_{-\infty}^\infty\mbox{d}x\, \delta\left(x-\sum_{i=0}^{n-1} \delta x_i\right)e^{ikx} = e^{ik\sum_{i=0}^{n-1} \delta x_i}$, %we can write $$\hat{P}(k,n)=\int_{-\infty}^\infty\mbox{d} x_{n-1}\int_{-\infty}^\infty \mbox{d} x_{n-2}...W(\delta x_{n-1})W(\delta x_{n-2})...W(\delta x_n)\exp(ik\sum_{i=1}^n x_i),$$ 
if the increments are also \textit{independent}, we can separate the integrals in Eq.~\eqref{CLT1} and write 
\begin{eqnarray} 
\hat{P}(k,n)=&\left[\int_{-\infty}^\infty\!\mbox{d} x_{n-1}W(\delta x_{n-1})e^{ik\delta x_{n-1}}\!\right] \nonumber\\
&\left[\int_{-\infty}^\infty \!\mbox{d} x_{n-2}W(\delta x_{n-2})e^{ik\delta x_{n-2}}\!\right]\cdots \nonumber\\ &\left[\int_{-\infty}^\infty\!\mbox{d} x_{0}W(\delta x_{0})e^{ik\delta x_{0}}\!\right]\!=\![\hat{W}(k)]^n.
\label{CLT2}
\end{eqnarray}
Finally, for a finite increment variance, it can be shown that in the limit $k\rightarrow 0$ (associated with large $\delta x_i$), one has  $\hat{W}(k)\simeq 1-\sigma^2k^2/2$~\cite{klafter2011first}.
%due to the Tauberian theorem
%+\mathcal{O}(k^\nu)$ ($2<\nu$ depends on the particular details of the system).  
Performing the inverse Fourier transform 
$\hat{f}(x)\rightarrow f(k)=\frac{1}{2\pi}\int_{-\infty}^\infty f(x)e^{-ikx}\mbox{d} k$,
we thus obtain $P(x,n)=\frac{1}{\sqrt{2\pi \sigma^2 n}}\exp[-x^2/(2n\sigma^2)]$, a Gaussian distribution, as expected. 

Violations of this derivation occur in the following scenarios:
(i) If the increment PDF in Eq.~\eqref{CLT1} explicitly depends on time (Moses effect), one has $W(\delta x_i)\rightarrow W(\delta x_i,i)$. Thus, the relation $\hat{P}(k,n)=[\hat{W}(k)]^n$ is no longer valid, since  $\hat{W}(k)\rightarrow\hat{W}(k,i)$ depends on $i$, and this may lead to  time dependence in the product $\hat{W}(k,0)\hat{W}(k,1)...\hat{W}(k,n-1)$.  
(ii) In the presence of temporal autocorrelations (Joseph effect), the integrals cannot be separated, rendering Eq.~\eqref{CLT2} invalid. 
(iii) If the variance of the increments is infinite (Noah effect), the asymptotic shape of $\hat{W}(k)$ may include non-integer power-laws in $k$ yielding a nonlinear-in-time MSD. 

%%%%% 
%%%%% End of Derivation of CLT Appendix. 
 
\section{Evaluation of the Joseph exponent $J$ from the TAMSD} 
\label{SubsecJAndTASD}

%In Table $1$, 
The Joseph exponent $J$ is defined via the scaling of the integrated velocity-autocorrelation function, with respect to the time-gap between the two time points~\cite{aghion2021moses}  
\begin{equation}
    \int_{0}^{s} \frac{\average{\mathbf{v}(t)\cdot \mathbf{v}(t+s')}}{\average{\mathbf{v}^2(t)}}\mbox{d} s' \propto s^{2J - 1},  
    \label{AutoCorrelationsJosephDeff}
\end{equation}
for ${s}\in[s_c,\infty)$, and $s_c>0$ is some lower cutoff. 
%This can lead to anomalous diffusion if the sign behind the proportionality symbol is positive or negative and $J>1/2$, since the temporal correlations decay slowly with $s$ (i.e., they are long-ranged), or when $J<1/2$ and the sign is negative, due to the  anti-correlations  ~\cite{mandelbrot1968noah,beran2017statistics}. In~\cite{meyer2017greenkubo,meyer2018anomalous,aghion2021moses} this definition is also discussed in the context of small $s$ asymptotic.  
The autocorrelation function $\average{\mathbf{v}(t)\cdot \mathbf{v}(t+s')}$, however, is often difficult to measure directly from data, since it requires a very large ensemble of long trajectories to overcome the  noise. For this reason, several alternative numerical methods have been developed to measure this exponent from various other observables, that are mathematically linked to  Eq.~\eqref{AutoCorrelationsJosephDeff}, see e.g. Refs.~\cite{chen2017anomalous,WAVELET,Hoell19,ANDI2021}. 
%To name a few: In~\cite{Pen94,Hoell19}, it was shown that Eq. \eqref{AutoCorrelationsJosephDeff} is directly related to the so-called fluctuation-function  in \textit{detrended fluctuations analysis}, which has been used e.g., in~\cite{meyer2018anomalous,aghion2021moses,ANDI2021}. In~\cite{WAVELET}, it was shown that this fluctuations-function can also be evaluated using \textit{wavelet analysis}. In~\cite{Bas07,chen2017anomalous}, $J$ is evaluated using \textit{rescaled
%range statistics}. 

We chose to compute $J$ using Eq.~\eqref{eq4}, which is computationally inexpensive compared to the other techniques, and easy to implement. Here we generalize the derivation of the link between Eqs.~(\ref{AutoCorrelationsJosephDeff}) and \eqref{eq4}~\cite{aghion2021moses,meyer2017greenkubo} for  $d\geq1$ dimensions. 
We start from the TAMSD  
\begin{equation}
\label{TAMSDdivide}
\left\langle \overline{\delta^2(s,t)}\right\rangle \approx  \frac{1}{t-s}\!\int_0 ^{t-s} 
\!\left\langle \left[ \mathbf{x}(t'\!+\!s) \!-\! \mathbf{x}(t')\right]^2\right\rangle  \mbox{d} t'.
\end{equation}
Focusing on the long time limit and also assuming $t\gg s$, we can use the Green-Kubo relation to write~\cite{meyer2017greenkubo} 
\begin{equation}
    \left\langle\! \left[ \mathbf{x}(t'\!+\!s) \!-\! \mathbf{x}(t')\right]^2 \!\right\rangle \!=\!  2\!\int_0^{s}\!\!\! dt_2\!\int_0^{t_2}\!\!\! dt_1 \!\left\langle \mathbf{v}(t_1\!+\!t')\!\cdot\!\mathbf{v}(t_2\!+\!t') \right\rangle\!.
    \label{xmxofC}
\end{equation} 
Equations~(\ref{TAMSDdivide},~\ref{xmxofC}) allow tying between the asymptotic scaling shape of the TAMSD and the autocorrelation function, given by Eq.~\eqref{AutoCorrelationsJosephDeff}. The crux of the derivations in~\cite{aghion2021moses,meyer2017greenkubo} is to write the autocorrelation function in a  general scaling form, depending on the properties of the process, which eventually lead to $\langle\overline{\delta(s,t)}\rangle\sim t^{2M+2L-1}s^{2J}$. This scaling, although different from that of the autocorrelation function, allows finding the exponent $J$ in an independent manner from $M$ and $L$, see main text. Note that, these derivations were originally done in one dimension, but can be easily extended to the scalar product $\langle \mathbf{v}(t_1+t')\cdot\mathbf{v}(t_2+t')\rangle$. In addition, the details of the derivation depend on the properties of the autocorrelation function, e.g., whether $J$ is above or below $1/2$. 

%Although the TAMSD is a very convenient observable, both from the interpretation point of view and measurement perspective, when comparing between the sum of the exponents $J+M+L-1$ and $H$ (see Fig. 5), our numerical results do not fully restore  the summation relation in Eq.~\eqref{summation_relation}. Instead, a small difference of $\sim4\%$ on average is observed. To study this discrepancy, we have compared between the predictions for $J$ of different methods, obtained from  ensembles of simulated fractional Brownian motion paths with various values of $J$. Based on our results (not shown), we find that extracting $J$ from the TAMSD gives rise to a slight underestimation of the true value of the Joseph exponent, when the ensemble size is small and/or the trajectories are short. In this case, one may use other methods to estimate $J$, such as wavelets~\cite{WAVELET}. 
%However, as the length of the trajectories or ensemble size are increased, we checked that the values of $J$ converge to the true known value of the simulation. 

\section{Statistical analysis}
\label{SubSecLocalScaling}
    \textbf{Local scaling}.
    In several data sets, the scaling regimes fitted to a power-law are local and do not span orders of magnitude. Nonetheless, in all cases reported here, a local scaling exponent can be fitted to the data, in discernible regimes. As the scaling for all four of our empirical quantities is tied through the summation relation, we view the local exponents as a biologically/physically  meaningful scaling. Naturally, in few cases, fitting the data in other temporal regimes may reveal slightly different scaling exponents. Yet, we expect all such power-law scaling to maintain the summation relation and to hold significant  information at the fitted scale.  
    
    \textbf{Error analysis}.
    For all data sets the fits were performed using SciPy library's curve-fit (nonlinear least squares method) in python 3.8. In order to satisfy the physical constraint of $L \geq 0.50$ we added bounds to the fits on the mean absolute velocity and mean-squared velocity such that this physical constraint is satisfied. \iffalse (we checked that the results without the constraint were the same within the measurement error) \fi As an initial error estimate we took one standard deviation for the parameter of the fitted power-law exponent (error 1). As the local regimes were visually identified, another source of error can be the number of points included in a fitted regime (error 2). To quantify this error we repeated the fit after excluding 5\% of data points on the sides of the corresponding regime, and computed the difference between the fits when including these points and when excluding them. In cases where removing the points on the regime boundaries led to large errors ($>10\%$ of a measured exponent) we deduced that no local exponent exist. For instance, when measuring $J$, denoting by $\Delta J_{1}$ and $\Delta J_{2}$ error 1 and 2 in $J$, the total error was $\Delta J = \sqrt{\Delta J_{1}^2 + \Delta J_{2}^2}$. Note, that this is taken as the upper bound on the error, since these two sources of error can be correlated.  
    The total error on the predicted value $H_p$ is thus given by $\Delta H_{p} = \sqrt{\Delta J^2 + \Delta M^2 + \Delta L^2}$. This formula gives an upper bound for the error on H, as it assumes that the errors on the exponents are independent, which is not necessarily the case for these data. 
    
    \textbf{Missing data points.}
    In several data sets it is common to encounter missing data points in a time series~\cite{fleming2014fine}. Discarding such time series from the ensemble is possible, yet undesirable. Hence, for any time-series that has $>$90\% of the points present, we fill each missing data point with a NaN (Not a Number), \textit{i.e.}, an empty placeholder which is naturally not included in the averaging. To verify that this choice does not affect the statistics, we checked that our results do not change when varying the 90\% threshold between 70\% and 95\%. Furthermore, we checked that in CTRW simulations, randomly replacing 10\% of the data points with NaNs does not significantly affect the results.
    %Note that here, we do not consider unevenly sampled data; yet, we refer the reader to Ref.~\cite{fleming2014fine}, in which calculations of the EA TAMSD was performed for unevenly sampled data. 
    
    \textbf{Multidimensional data}.
    Previous works on the Joseph, Moses and Noah effect treated only unidimensional simulations~\cite{chen2017anomalous,meyer2018anomalous, aghion2021moses}. Here, we expand the framework to multidimensional data, by reducing two dimensional $[x(t),y(t)]$ to one dimensional time series $|\mathbf{x}(t)|$ (Fig.~\ref{fig:Scheme}). In general, it is not trivial that such a projection will yield results that are similar to any of the original $x(t)$ or $y(t)$. Thus, in the cases studied here we verified that averaging over any of the original variables gives similar results to averaging over their projection.

\textbf{Measuring the exponents.} 
\label{AppenSec1_3}
Here we provide a recipe to generate a time-series from raw empirical data, and obtain the exponents $M, L$ and $J$ for a path ensemble. 
\begin{enumerate}
    \item 
    We choose a constant sampling frequency, to generate uniformly sampled time-series from the raw data sets. If a small percentage of the time series is missing in the data ($<$ 10\%), the missing locations are treated as NaNs and are excluded from any averaging (see above). 
    \item 
    For each trajectory $\mathbf{x}(t)$ in $d\geq1$ dimensions, where $t$ is the total measurement time, we choose an additional constant time increment of duration $0<\Delta\ll t$. The size of $\Delta$ should be larger than the sampling rate of the data, but much smaller than the total duration of the time series. For different values of $\Delta\ll t$, we obtain
    \begin{enumerate}
        
        \item 
        Time-averaged absolute velocity 
{        \begin{equation} \label{MeanAbs}
            \overline{|\mathbf{v}|}(t) \equiv \frac{\Delta}{t}\sum_{j = 1}^{t/\Delta} \frac{|\delta \mathbf{x}_j|}{\Delta} 
        \end{equation}}
        where  $\delta \mathbf{x}_j \equiv \mathbf{x}(j\Delta) - \mathbf{x}([j-1]\Delta)$ is the $j$th vector-increment of the path. 
        %Note that $\delta\mathbf{x}_j$ generalizes this definition for multidimensional time series. 
        \item 
        Time-averaged squared velocity 
        {\begin{equation}
           \label{MeanSqr} \overline{\mathbf{v}^2}(t) \equiv \frac{\Delta}{t}\sum_{j = 1}^{t/\Delta} \frac{(|\delta \mathbf{x}_j|)^2}{\Delta^2}. 
        \end{equation}}
    \end{enumerate}
    \item 
    For each trajectory,  compute the time averaged squared displacement  
        \begin{equation} \label{TASD}
            \tasd \equiv \frac{1}{t - s}\int_0^{t - s} [\bm{x}(t' + s) - \bm{x}(t')]^2 dt',
        \end{equation}
        as a function of $s$. 
    \item 
    After repeating steps 2 and 3 for all the time series, we can compute the ensemble average of the quantities above, for each time $t$ [in the case of Eqs. \eqref{MeanAbs} and \eqref{MeanSqr}], and each value of $s\in[\Delta, t]$ [in the case of Eq.~\eqref{TASD}] over all time-series in the ensemble.
    \item 
    By plotting the statistics (see main text and Sec.~S1 in~\cite{SM}) we visually identify regimes that can be described by local scaling exponents. We then fit each regime with a power law using the method of non-linear least squares. In order to satisfy the physical constraint of $L \geq 0.50$ we add this bound by first fitting $\langle\overline{\mathbf{v}^2}\rangle$ to a scaling exponent, and then constraining the fit of $\langle\overline{|\mathbf{v}|}\rangle$ in order to satisfy Eq.~(2) of the main text under the condition that $L \geq 0.5$. Note that, when applying  power-law fits to data, one may consider to  apply additional statistical tests to gain confidence about the results of the local exponents. 
\end{enumerate}

\section{Continuous-time random walk simulations}
\label{SubSecCTRW}
In Table~\ref{tab:exponents}, in addition to the experimental results, we added results of simulations, which were performed for multiple dynamical models. These models are not intended to fully explain the dynamics of the
experimental systems; rather they can provide valuable insights into the relations between the various exponents. Simulations for several prototypical examples are cited from Ref.~\cite{chen2017anomalous}, whereas CTRW simulations, with an asymptotic power-law waiting-time distribution (see below), for free, bounded and biased random walkers were performed as part of the current study (see, e.g., Fig. S13 in \cite{SM}). 

CTRW is a random walk defined in terms of the waiting time $\tau$ between successive jumps -- a random variable drawn from the PDF $\psi(\tau)$. When the average waiting time $\average{\tau}$ diverges, the process displays subdiffusive dynamics, weak ergodicity breaking and aging~\cite{metzler2014anomalous}. 
% CTRW were used in previous works to model quantum dots~\cite{brokmann2003statistical, margolin2005nonergodicity}, mRNPs~\cite{song2018neuronal} and search patterns of avian predators~\cite{vilk2021ergodicity}. Notably, the summation relation, Eq. \eqref{summation_relation}, has not been validated for CTRW in previous works. 
In accordance with empirical data, see \textit{e.g.} Ref.~\cite{brokmann2003statistical,song2018neuronal,
vilk2021ergodicity}, we assume power-law distributed waiting times, $\;\psi(\tau) \sim \tau^{-(1+\alpha)}$, for $0<\alpha<1$. We simulated three cases: free CTRW, bounded CTRW (BoCTRW), and biased CTRW (BiCTRW) for different values of $\alpha$. In BoCTRW the random walker is bounded by a confining potential~\cite{metzler2014anomalous}, in the sense that it cannot exit predefined domain walls but can move freely within these walls. In BiCTRW the turning angle of each jump (in radians) is sampled from a wrapped Cauchy distribution defined by 
$f(x) = (1 - \rho^2)/[2\pi (1 + \rho^2 - 2 \rho \cos x)]$ 
with $\rho<1$. In Table~\ref{tab:exponents} we simulated the case of $\rho = 0.3$. The results for BiCTRW match the theoretical results in~\cite{bel2005weak}. 
For all processes we simulated an ensemble of 1000 trajectories of length $t = 1500$ time steps. 
Notably, for both free and bounded CTRW we obtain similar values of $L$ and $M$, but for the latter $J$ decreases dramatically due to boundary interactions~\cite{metzler2014anomalous}, indicating a transition from positive long-ranged correlations, to anti-correlated motion. Also note that CTRW couples between $M$ and $L$. For $\alpha<1$, a negative Moses effect arises due to increasingly long waiting times experienced by the particle as time evolves, which slow down the dynamics. The Noah effect emerges since at most times, the random-walker is stuck in a single location, and practically any jump is a rare  event. As  $\alpha$ approaches $1$, the waiting times become shorter, and jumps become more frequent (on average); hence these effects vanish, and $M$ and $L$ approach $1/2$ (for $\alpha>1$, $\average{\tau}$ becomes finite, and there is no longer significant  aging~\cite{bel2005weak,burov2011single}).

\section{p-variation test} \label{appendix:p}
% Here we employ the p-variation test (see below) to further validate our claim that the movement within ARS is adequately modeled by subdiffusive CTRW. 
For several of the processes detailed above we performed a p-variation test to distinguish non-Gaussian CTRW from other types of subdiffusive behaviors such as the Gaussian FBM~\cite{metzler2014anomalous, magdziarz2010detecting, magdziarz2009fractional}, see also discussion in \cite{jeon2013noisy}.
The test is defined in terms of the sum of increments of a trajectory $x(t)$ on the time interval $[0, T]$:
\begin{equation}
V_n^{(p)}(t) \!=\!\! \sum_{j = 0}^{2^n\! -\! 1}  \left|  x\!\left( \!\min\! \left\{\! \frac{(j\!+\!1)T}{2^n}, t\right\}  \!\right) \! -\! x\!\left(\!\min\! \left\{ \!\frac{j T}{2^n}, t\right\}\!  \right)  \right|^p\!\!. 
\end{equation}
For FBM $V^{(p)}(t) = \lim_{n\to \infty} V_n^{(p)}(t)$ displays the following properties: for $p = 1/H$ it tends to be linear with the observation time $t$, while for $p > 1/H$ it is equal zero and for $p < 1/H$ it is equal to infinity~\cite{magdziarz2009fractional}. 
In contrast, for subdiffusive CTRW, $V^{(p)}(t) = \lim_{n\to \infty} V_n^{(p)}(t)$; for $p = 2$ it shows a monotonic, step-like increase in time, while for $p=2/\alpha$: $V^{(2/\alpha)}(t) = 0$~\cite{magdziarz2009fractional, magdziarz2010detecting}, $\alpha$ being the parameter for the CTRW, see Appendix \ref{SubSecCTRW} above. 
In Sec.~S4 and Fig.~S14 in~\cite{SM}, we show an example of this test on randomly chosen amoeba tracks, and the test shows good agreement with the suggested FBM-like dynamics detailed above. 
In contrast, for the searching kite (Fig.~S15 in~\cite{SM}) the test suggests CTRW dynamics with $\alpha = 0.5$ (as obtained in Ref. \cite{vilk2021ergodicity}). 
For the stork we perform the test during all seasons and observe good agreement with the models suggested above. For instance, trajectories during breeding (June-July) show clear characteristics of CTRW (Fig. S16), while trajectories during migration (e.g., September) show remarkable agreement with the theory for FBM (Fig. S17). 
Notably, in each of the above cases we have repeated the test on many randomly chosen trajectories.

\bibliography{scibib}

\clearpage
\renewcommand{\thefigure}{S\arabic{figure}}    
\setcounter{figure}{0} 

\renewcommand{\thesection}{S\arabic{figure}}    
\setcounter{figure}{0} 

\onecolumngrid
\section*{Supplemental Material} 

Here we present statistics for data sets that are not presented in the main text. All exponents that result from these statistics are detailed in Table 1 of the main text. 
Below, the notations and acronyms are the same as in the main text and the equations and figures refer to those therein.
 
\section*{S1: Data analysis for all datasets} \label{sec:ResultsSI}

\subsection{Fluorescent Rhodamine molecules}
\label{Appen2_1}Here we plot the statistics for the fluorescent Rhodamine molecules in different relative humidity (see main text). For all humidities, the trajectories are two dimensional and are acquired at an acquisition time of $\Delta = 50$ ms, for at least 1500 ms (some trajectories are longer). For the values of the exponents, see Table 1 in the main text. 

\begin{figure*}[h!]
\centering
\includegraphics[width=0.8\textwidth]{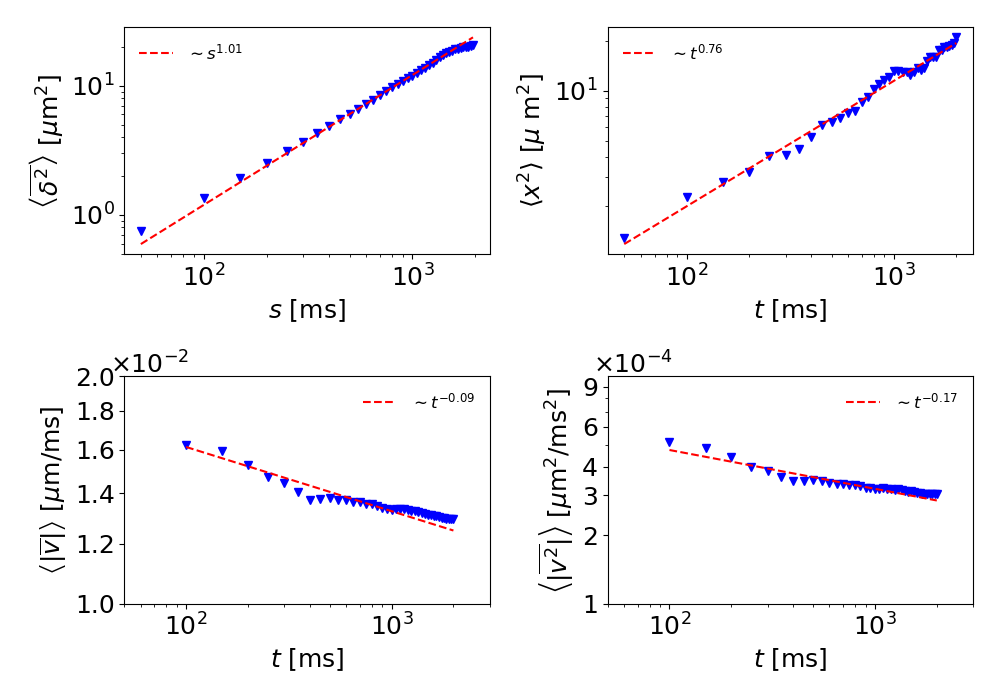}
 \caption {\footnotesize{Statistics for fluorescent Rhodamine molecules with relative humidity of 100\%. All panels were fitted in the range $50 < t < 1500$ ms. Measured values are plotted in blue triangles and fits as red dashed lines. %(a) The ensemble-averaged TASD scales as $s^{0.95}$ and $s^{0.97}$. (b) The MSD scales as $t^{0.71}$ and $t^{0.91}$. (c) Ensemble-averaged time-averaged velocity scales as $t^{-0.14}$ and $t^{-0.05}$. (d) Ensemble-averaged time-averaged squared velocity scales as $t^{-0.29}$ and $t^{-0.09}$.
 }}
 \label{fig:MolRH100}
\end{figure*}

\begin{figure*}[h!]
\centering
\includegraphics[width=0.8\textwidth]{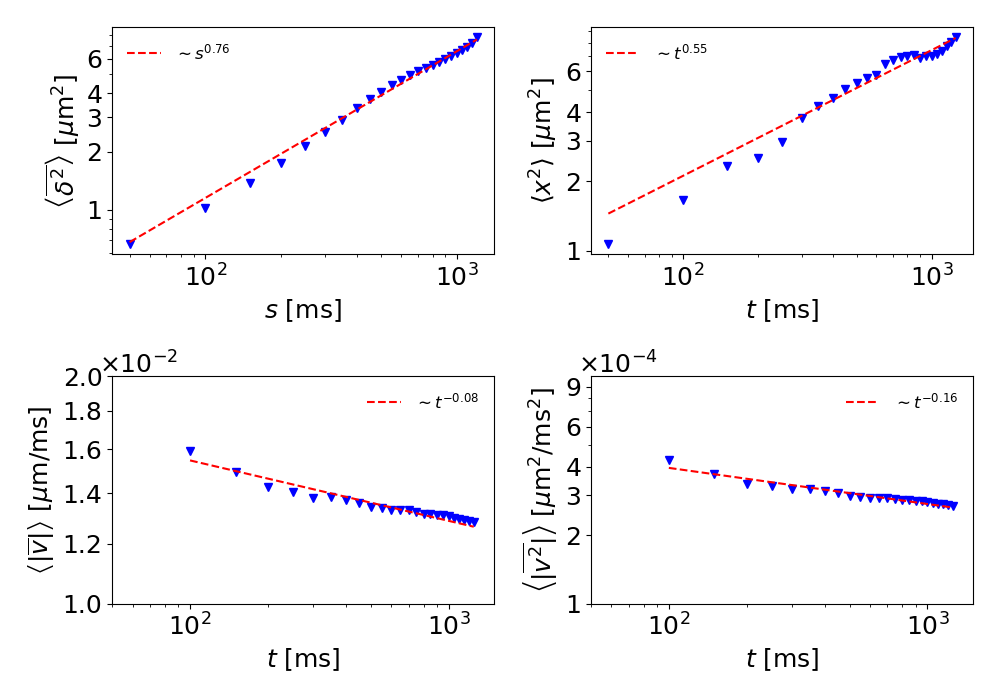}
 \caption {\footnotesize{Statistics for fluorescent Rhodamine molecules with relative humidity of 90\%. All panels were fitted for $50 < t  < 1500$ ms.  Measured values are plotted in blue triangles and fits as red dashed lines. %(a) The ensemble-averaged TASD scales as $s^{0.89}$. (b) The MSD scales as $t^{0.63}$. (c) Ensemble-averaged time-averaged velocity scales as $t^{-0.07}$. (d) Ensemble-averaged time-averaged squared velocity scales as $t^{-0.14}$.
 }}
 \label{fig:MolRH90}
\end{figure*}

\begin{figure*}[h!]
\centering
\includegraphics[width=0.8\textwidth]{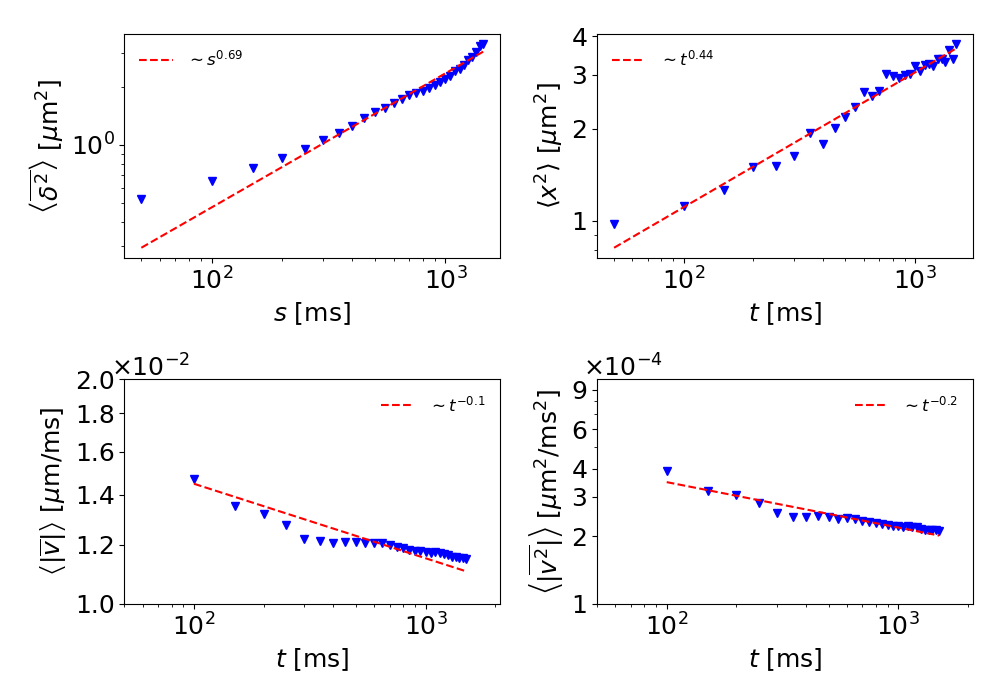}
 \caption {\footnotesize{Statistics for fluorescent Rhodamine molecules with relative humidity of 85\%. All panels were fitted for $50 < t  <1500$ ms.  Measured values are plotted in blue triangles and  fits as red dashed lines. %(a) The ensemble-averaged TASD scales as $s^{0.69}$. (b) The MSD scales as $t^{0.46}$. (c) Ensemble-averaged time-averaged velocity scales as $t^{-0.08}$. (d) Ensemble-averaged time-averaged squared velocity scales as $t^{-0.17}$.
 }}
 \label{fig:MolRH85}
\end{figure*}

\begin{figure*}[h!]
\centering
\includegraphics[width=0.8\textwidth]{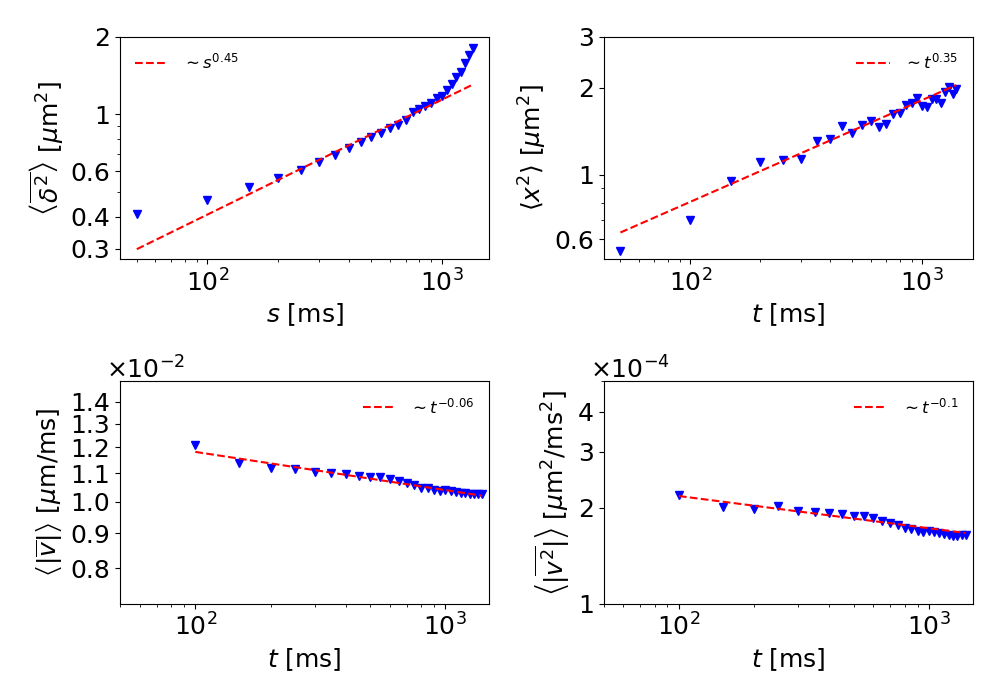}
 \caption {\footnotesize{Statistics for fluorescent Rhodamine molecules with relative humidity of 75\%. All panels were fitted for $50 < t  < 1500$ ms.  Measured values are plotted in blue triangles and  fits as red dashed lines. %(a) The ensemble-averaged TASD scales as $s^{0.47}$. (b) The MSD scales as $t^{0.33}$. (c) Ensemble-averaged time-averaged velocity scales as $t^{-0.05}$. (d) Ensemble-averaged time-averaged squared velocity scales as $t^{-0.11}$.
 }}
 \label{fig:MolRH75}
\end{figure*}

\begin{figure*}[h!]
\centering
\includegraphics[width=0.8\textwidth]{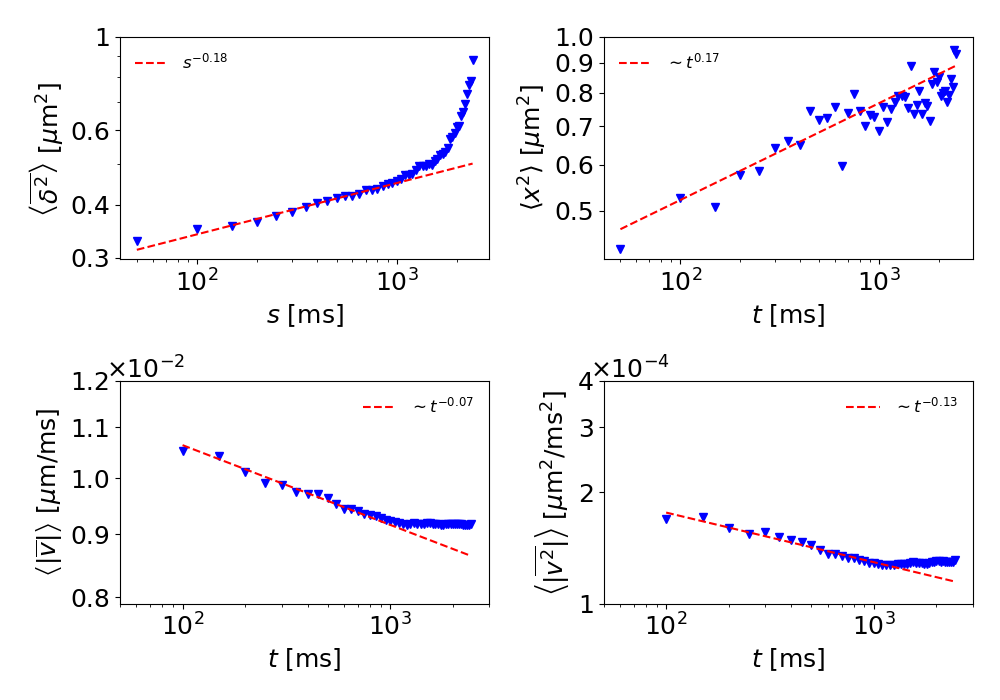}
 \caption {\footnotesize{Statistics for fluorescent Rhodamine molecules with relative humidity of 30\%. All panels were fitted for $50 < t  < 1500$ ms.  Measured values are plotted in blue triangles and  fits as red dashed lines. %(a) The ensemble-averaged TASD scales as $s^{0.19}$. (b) The MSD scales as $t^{0.14}$. (c) Ensemble-averaged time-averaged velocity scales as $t^{-0.06}$. (d) Ensemble-averaged time-averaged squared velocity scales as $t^{-0.11}$.
 }}
 \label{fig:MolRH30}
\end{figure*}

In Figs.~\ref{fig:MolRH100},~\ref{fig:MolRH90},~\ref{fig:MolRH85},~\ref{fig:MolRH75} and~\ref{fig:MolRH30} we plot the statistics for relative humidity of 100\%, 90\%, 85\%, 75\% and 30\%, where the statistics are averaged over an ensemble of 174, 298, 239, 258 and 436 trajectories, respectively.

%\clearpage

%\clearpage

%\clearpage

%\clearpage

% \subsection{Telomeres}
% In Fig.~\ref{fig:telomeres} we plot the statistics for the telomeres. Here the statistics are averaged over an ensemble of 20 two dimensional trajectories, with 1000 observations each, tracked with a time lag $\Delta = 200$ ms. In the ensemble-averaged TASD we clearly see two regimes, where for short times ($t < 6$) the dynamics are anti-correlated with $J = 0.31$ and for long times ($t > 10$) the dynamics are only very weakly anti-correlated with $J = 0.46$. In both cases the Moses and Noah effects are negligible, as the slopes of the ensemble-averaged time-averaged velocity and squared velocity are both close to zero. Note that the MSD cannot be fitted to a power law in any single discernible regime. 
% Our results are consistent with Ref.~\cite{stadler2017non}. 

% \begin{figure*}[h!]
% \centering
% \includegraphics[width=0.85\textwidth]{telomeres_exponent_fits.png}
%  \caption {\footnotesize{Statistics for telomeres. All panels were fitted in two regimes, $0.125 < t < 10$ s and $40 < t < 200$ s. The measured values are plotted in blue triangles and the fits are plotted as red and black dashed lines for short and long times respectively. (a) The ensemble-averaged TASD scales as $s^{0.65}$ and $s^{0.93}$. (b) The MSD. Here, due to considerable noise, we were not able to obtain a proper fit. (c) Ensemble-averaged time-averaged velocity scales as $t^{0.02}$ and $t^{-0.03}$. (d) Ensemble-averaged time-averaged squared velocity scales as $t^{0.04}$ and $t^{-0.05}$.}}
%  \label{fig:telomeres}
% \end{figure*} 

 \clearpage

\subsection{Tracers in cells} 
\label{Appen2_2}
\begin{figure*}[h!]
\centering
\includegraphics[width=0.8\textwidth]{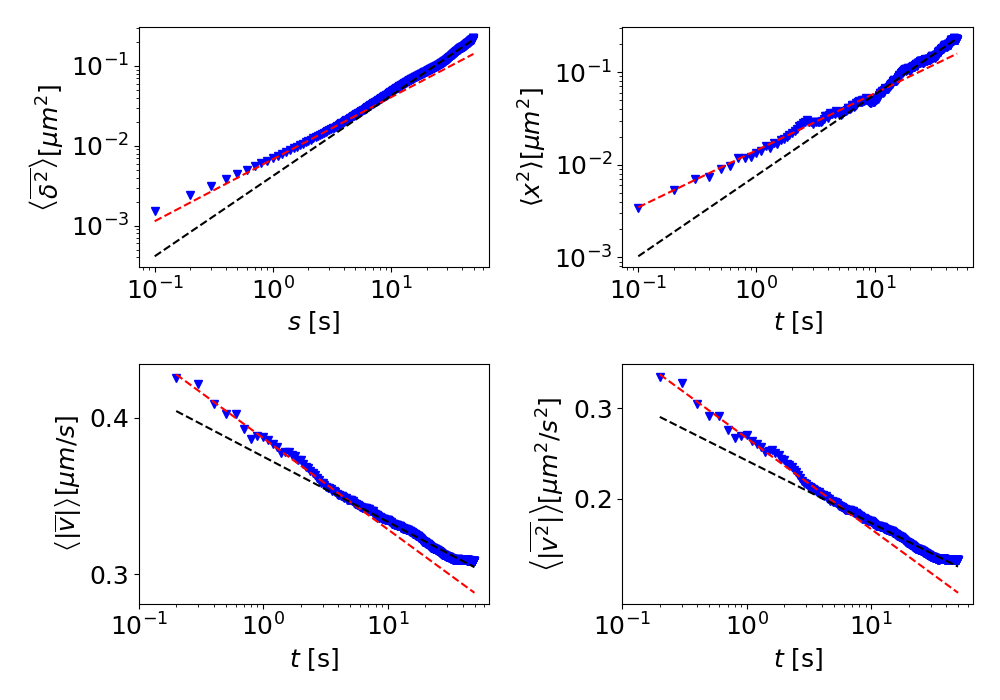}
 \caption {\footnotesize{ Tracers (quantum dots) in cytoplasm of treated mammalian cells.  All panels were fitted in two regimes, $0.1 < t < 5$ s and $5 < t < 50$ s. Measured values are plotted in blue triangles and fits as red and black dashed lines for short and long times respectively. }}
 \label{fig:qdots}
\end{figure*}

\begin{figure*}[ht!]
\centering
\includegraphics[width=0.8\textwidth]{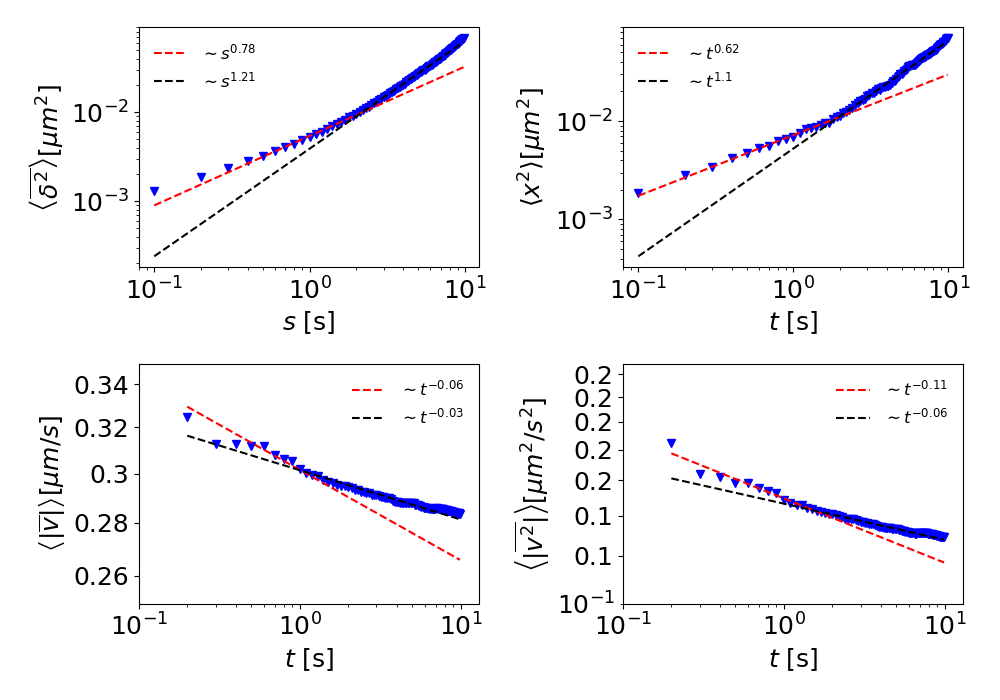}
 \caption {\footnotesize{ Tracers (quantum dots) in cytoplasm of untreated mammalian cells. All panels were fitted in two regimes, $0.1 < t < 2$ s and $2 < t < 8$ s. Measured values are plotted in blue triangles and fits as red and black dashed lines for short and long times respectively. }}
 \label{fig:qdots2}
\end{figure*}

In Fig.~\ref{fig:qdots} we plot the statistics for the tracers (quantum dots) in cytoplasm of treated mammalian cells (see main text). %Here the statistics are averaged over an ensemble of 200 one dimensional, 50 s long trajectories, sampled at $\Delta = 0.1 s$, see Ref.~\cite{sabri2020elucidating} for more details.
In Fig.~\ref{fig:qdots2} we plot the statistics for the untreated cells. %Here the statistics are averaged over an ensemble of 1000 one dimensional, 10 s long trajectories, sampled at $\Delta = 0.1 s$, see Ref.~\cite{sabri2020elucidating} for more details.
The results of the analysis (see Table 1 in main text) are consistent with the analyses performed in~[22]. 
There, it has been shown that the mean scaling exponent of TAMSDs (and their geometric ensemble average) for short time scales ($t<1s$) is very similar in both untreated and treated cells ($\alpha\approx0.58$, cf.~[22]). The average diffusion coefficient, however, was seen to be significantly lower for the ensemble of longer trajectories, highlighting a bias of long trajectories for lower mobilities. Strong fluctuations in the diffusion coefficient of individual trajectories also led to an overestimation of the short-term scaling exponent in both cases when using an arithmetic instead of a geometric averaging ($\alpha\approx0.78$, cf.~[22]). In fact, an arithmetic average is equivalent to the definition of Eq.~(4) of the main text, predicting a value $J=\alpha/2\approx 0.39$ for short time scales in both ensembles, in excellent agreement with our findings here. 
Fitting the arithmetic ensemble average of all TAMSDs on time scales $\tau>2$~s yielded an average scaling exponent $\alpha=1.12\pm0.07$ for untreated and $\alpha=0.90\pm0.02$ for latrunculin-treated cells, in good agreement with $J=0.60$ and $J=0.50$ in Table 1 of the main text.

\clearpage

\clearpage

\subsection{Harvester ants} 
\label{appen2_3}
\begin{figure*}[h!]
\centering
\includegraphics[width=0.8\textwidth]{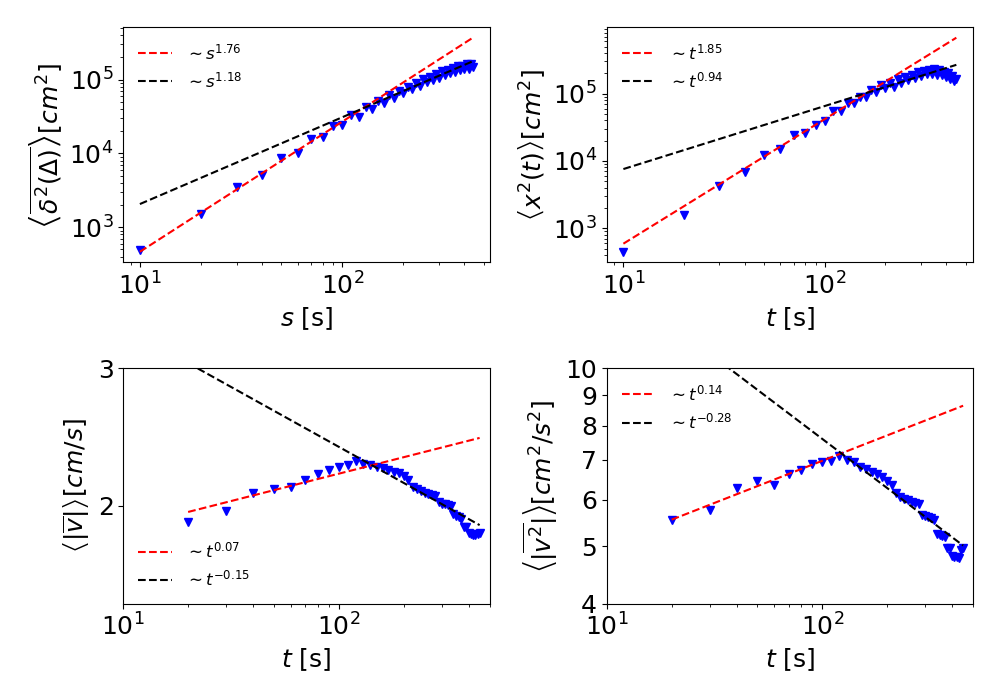}
 \caption {\footnotesize{ Harvester ants. All panels were fitted in two regimes, $10 < t < 100$ s and $100 < t < 400$ s. Measured values are plotted in blue triangles and fits as red and black dashed lines for short and long times respectively. }}
 \label{fig:ants}
\end{figure*}

In Fig.~\ref{fig:ants} we plot the statistics for the harvester ants. 

\clearpage

\subsection{Black winged kite} 
\label{Appen2_4}
\begin{figure*}[h!]
\centering
\includegraphics[width=0.8\textwidth]{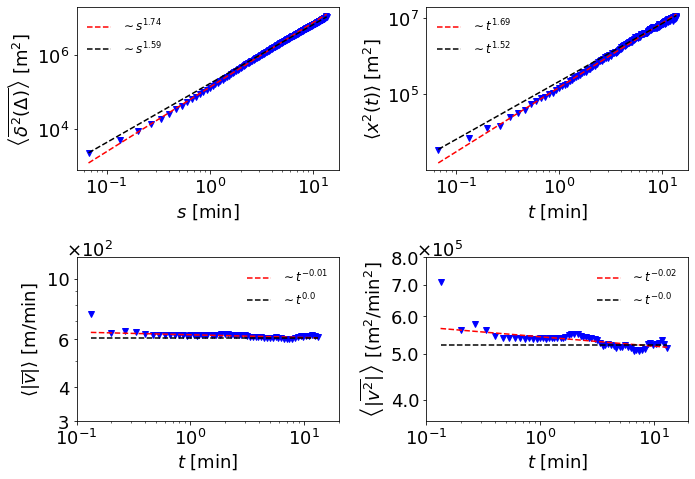}
 \caption {\footnotesize{kite -  commuting. All panels were fitted in two regimes: $0.1< t< 3$ and $3 < t < 12$ min. Measured values are plotted in blue triangles and fits as red (short times) and black (long times) dashed lines. Of these trajectories 107 last $\geq$ 10 min.}}
 \label{fig:kiteComm}
\end{figure*}

\begin{figure*}[h!]
\centering
\includegraphics[width=0.8\textwidth]{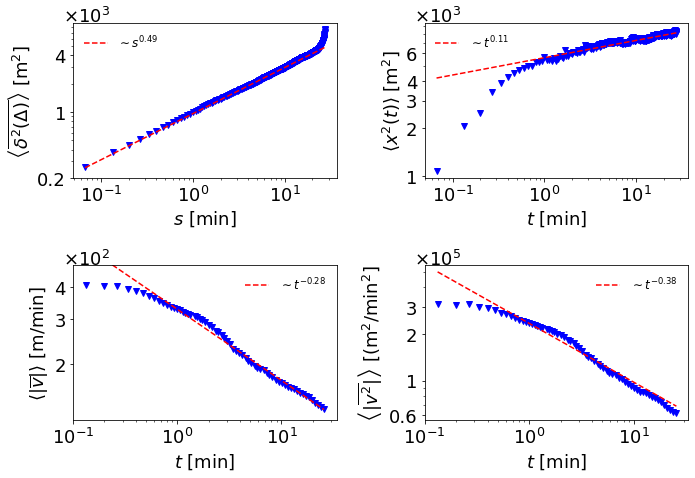}
 \caption {\footnotesize{kite searches. All panels were fitted for $0.1 < t < 20$ min. Measured values are plotted in blue triangles and fits as red dashed lines. Of these trajectories 587 last $\geq$ 27 min.}}
 \label{fig:kiteARS}
\end{figure*}

For the kite we independently analyzed the ensemble of commuting trajectories and area restricted searches, see main text. In Fig.~\ref{fig:kiteComm} and~\ref{fig:kiteARS} we respectively plot the statistics for the commuting and searching kite.%, averaged over an ensemble two-dimensional trajectories (recorded as locations in the Hula Valley local map), each measured at a frequency of 0.25 Hz. 

%Here  the statistics display two regimes, where for both short ($t < 2$ min) and long ($t > 4 $ min) the dynamics are positively correlated with $J = 0.88$ and $J = 0.77$ respectively. This matches the expected behavior during commuting flight, in which the animal covers large distances to get from one patch to another. Interestingly, for long times ($t > 4$ min) the correlation is smaller, entailing that at these times the trajectory is not as directed as for short time. This might reflect the fact that some flight trajectories (in particular the long ones) may be long-range exploratory modes, which are probably less correlated at long times. Here, the Moses and Noah effects are negligible. Thus, as the MSD and the EA TASD scale similarly, the process is ergodic~\cite{metzler2014anomalous, vilk2021ergodicity}. 

%\clearpage

%Here the statistics display a single regime in the panels of Fig.~\ref{fig:kiteARS}. The dynamics are anti-correlated with $J = 0.24$, which suggests that the animal is performing a search that is spatially restricted. Additionally, we find a measurable Moses and Noah effects, $M = 0.23$ and $L = 0.59$, which could be caused by long waiting times as found in Ref.~\cite{vilk2021ergodicity}, see the CTRW below. 

\clearpage

\subsection{White stork} 
In Fig. \ref{fig:storkGPSplot} we plot the statistics for the stork during wintering (A) and the spring migration (B). 

\begin{figure*}[h!]
\centering
\includegraphics[width=1.0\linewidth]{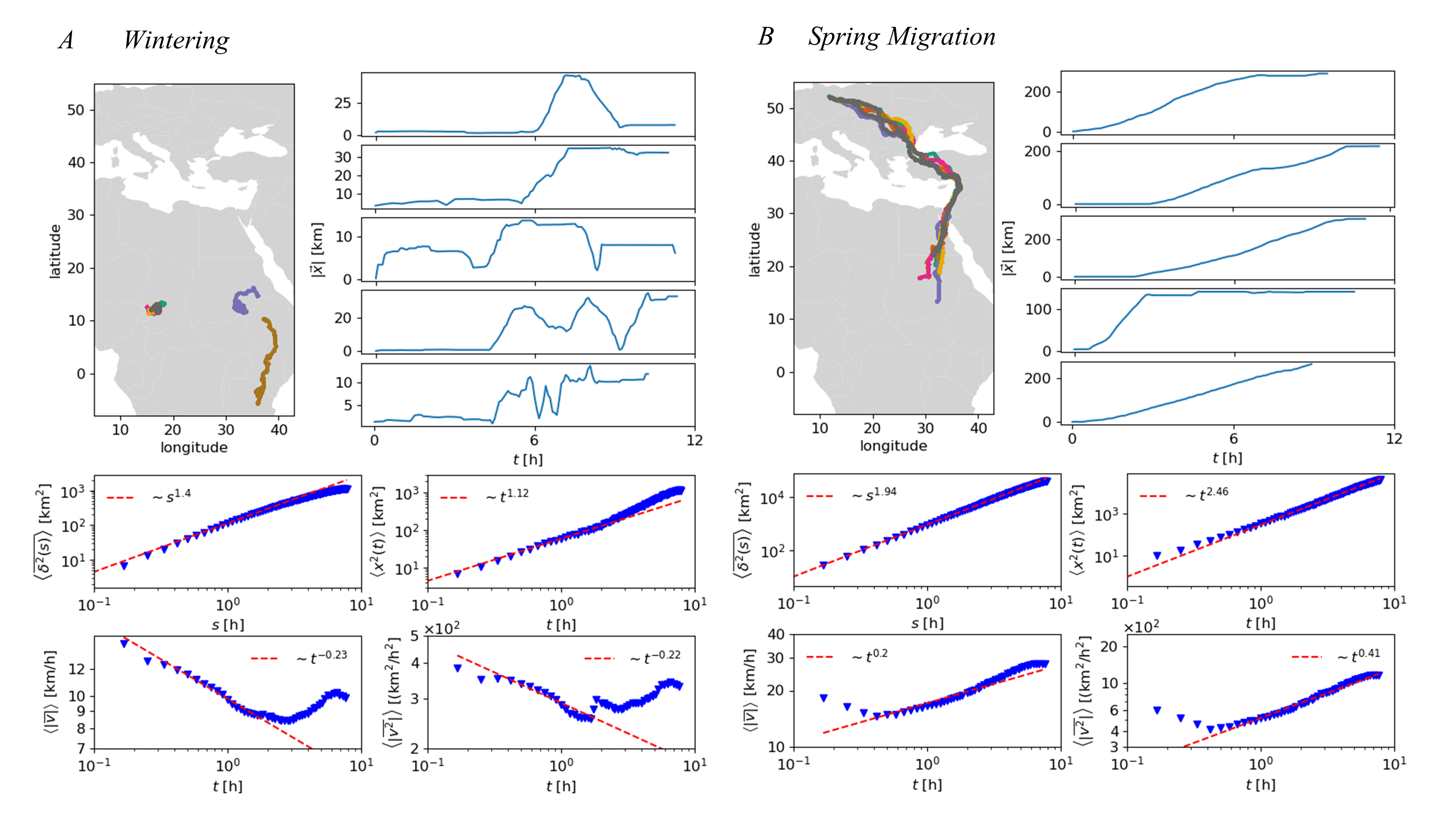}
 \caption {\footnotesize{Stork: from raw data to statistics during October-January (A, wintering) and March-April (B, spring migration). In each of A and B, the upper left panels are GPS tracks, where different colors represent different years. The five upper right panels are examples for the distance traveled at a single day as a function of the time of measurement in hours. The four lower panels are the statistics on the ensemble of days during the relevant period, and the fit values are given in the plot legends (see Fig. 3 for details). %Notably, in A the MSD could not be fitted to a power-law, due to considerable noise, yet the value of $H$ can still be estimated from the other exponents using Eq.~\eqref{summation_relation}.
 }}
 \label{fig:storkGPSplot}
\end{figure*} 

\clearpage
\section*{S2: Ageing and the Moses effect} 
\label{Appen3}
\begin{figure*}[h!]
\centering
\includegraphics[width=0.8\textwidth]{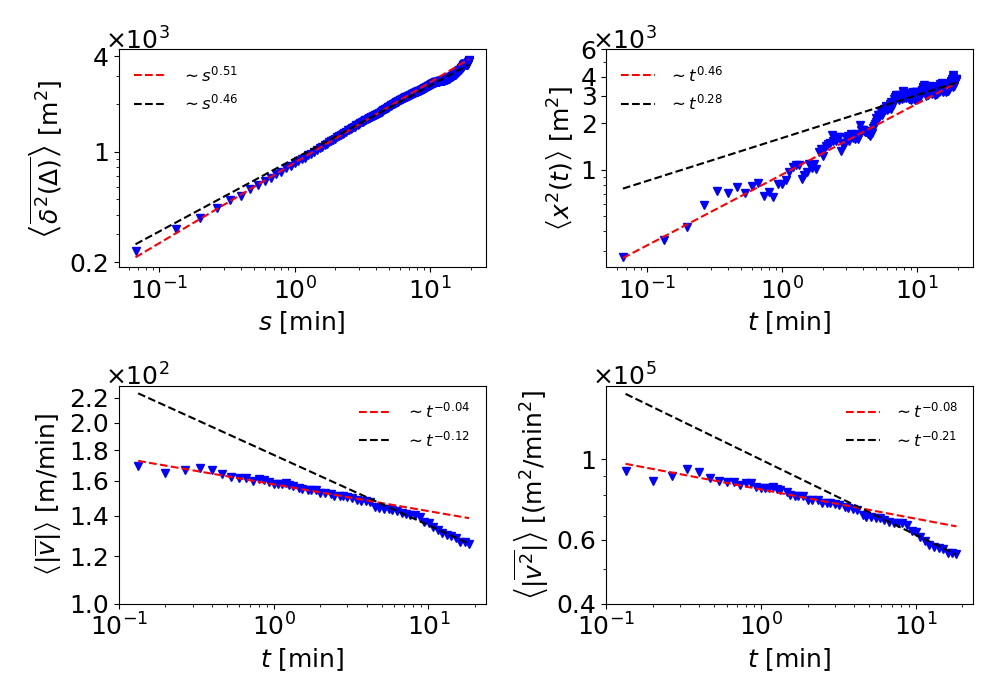}
 \caption {\footnotesize{kite searches - aged system. All panels were fitted in two regimes, for $0.1 < t < 5$ min and $5 < t < 20$ min.  Measured values are plotted in blue triangles and fits as red dashed lines.}}
 \label{fig:kiteARSaged}
\end{figure*}

In data sets with a measurable Moses effect, it is plausible that the data set also shows statistical ageing (see main text). Thus, the time of start of measurement can play a crucial role in determining the statistics of an experimental system~[7, 70]. 
To show this we use the ensemble of kite search flights described in the main text, and we age the system by "starting" each trajectory 10 min after first recorded measurement of that search. In Fig.~\ref{fig:kiteARSaged} we plot the statistics for this aged ensemble. The parameters now vary between two temporal regimes, and assume different values than those we find for the unaged system. Here, for the aged system, at times  $0.1 < t < 5$ min we find $J = 0.26$, $M = 0.46$, $L = 0.50$ and $H = 0.23$ (the summation relation predicts $H = 0.22$) and  at times  $5 < t < 20$ min we find $J = 0.23$, $M = 0.38$, $L = 0.52$ and $H = 0.14$ (the summation relation predicts $H = 0.12$).
As expected, the long range correlations are not strongly affected by the ageing~[7]. 
However, the Moses and Noah effects are significantly obscured by ageing, although the summation relation still gives a valid prediction.
These results entail that in order to fully capture a systems statistics, initiating the measurement at the beginning of the process can prove crucial. Ecologically, to describe any non-ergodic behavioral mode, one has to have the ability to measure the process from its beginning (this is both a conceptual challenge and a technical one, but it underlies our ability to get meaningful results from ecological tracking). 

% \subsubsection{Issue 2: large variability in the ensemble}

\clearpage

\section*{S3: Continuous-time random walk simulations} 
\label{Appen4}
\begin{figure*}[h!]
\centering
\includegraphics[width=0.8\textwidth]{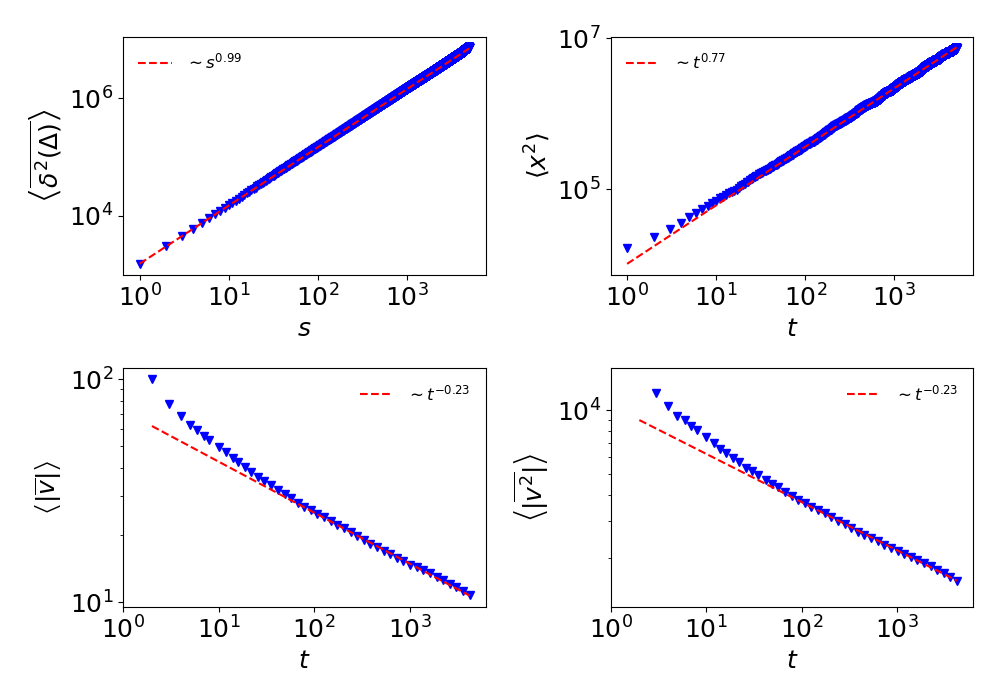}
 \caption {\footnotesize{Free CTRW with $\alpha = 0.8$. All panels were fitted for $100 < t < 5000$. Measured values are plotted in blue triangles and fits as dashed lines.}}
 \label{fig:CTRWalpha04Free}
\end{figure*}

In the main text we present results from CTRW simulations. Here we provide an example for statistics obtained from one simulation. In Fig.~\ref{fig:CTRWalpha04Free} we plot the statistics for free CTRW for $\alpha = 0.4$. Here we average 1000 simulations of 5000 points. The values for the exponents are provided in Table 1 of the main text.

\clearpage

\section*{S4: Supplementary analysis: p-variation test} 
\begin{figure}[h!]
	\center
	\includegraphics[width=.9\linewidth]{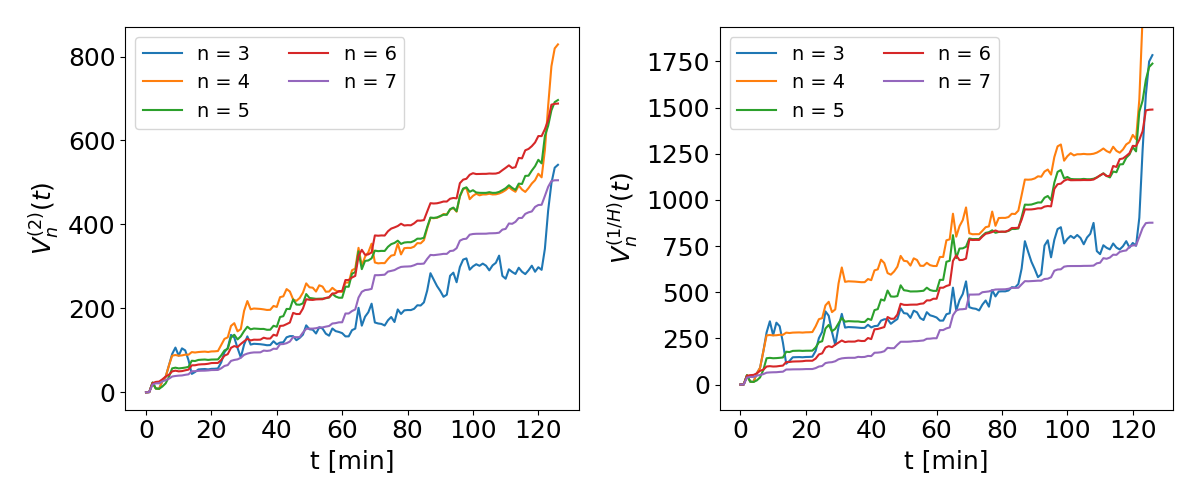}
	\caption{A p-variation test on a randomly chosen amoeba track. On the left panel $p = 2$, and on the right panel $p = 1/H$ for $H = 0.4$. Here the trends resemble those of Gaussian FBM~[61], as suggested in the main text and in agreement with the analysis performed in~[31]. Note that $n$ can only be increased up to $2^n = N$, $N$ being the number of data points in the trajectory. }
	\label{fig:pvar_amoeba}
\end{figure}

\begin{figure}[h!]
	\center
	\includegraphics[width=.9\linewidth]{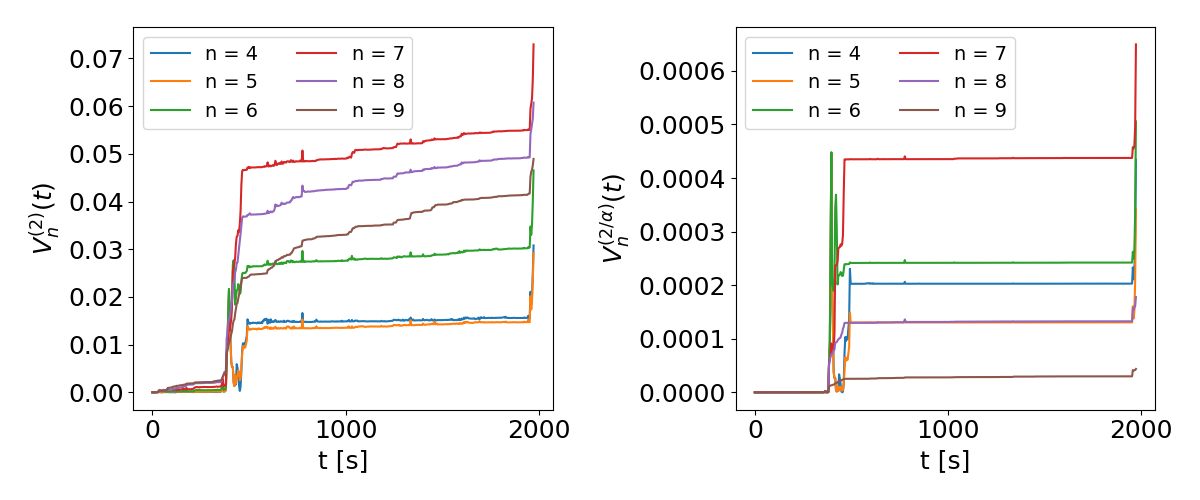}
	\caption{A p-variation test on a randomly chosen search segment of the kite. On the left panel $p = 2$, and $ V_n^{(p)}(t)$ displays a monotonic step-like increase. In (b) shown is the test for $p = 2/\alpha$ for $\alpha = 0.5$, and as expected $V_n^{(p)}(t)$ tends to zero as $n$ is increased.}
	\label{fig:pvar_kite}
\end{figure}

\begin{figure}[h!]
	\center
	\includegraphics[width=.9\linewidth]{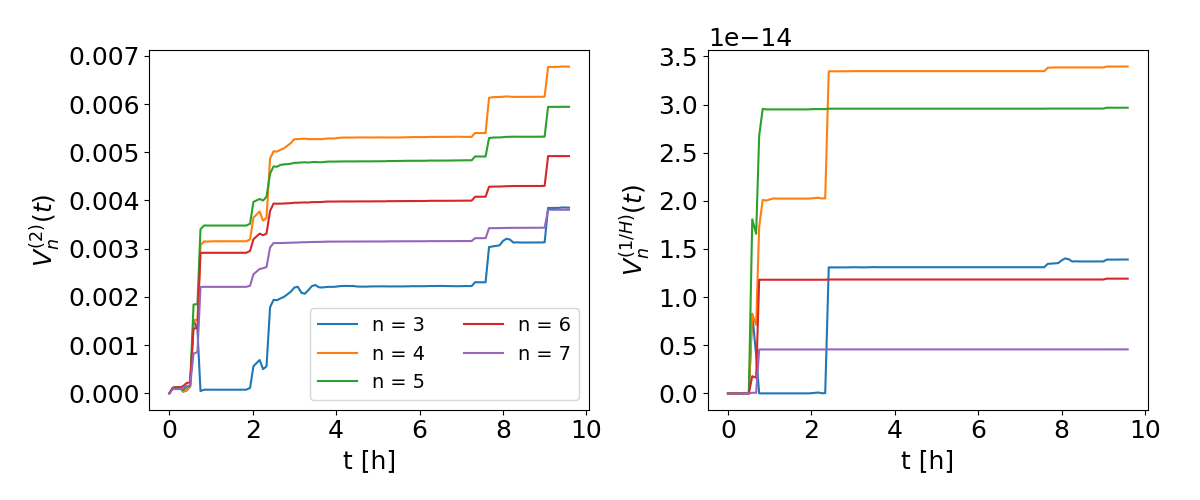}
	\caption{A p-variation test on a randomly chosen daily segment of the breeding stork (June). In the left panel $p = 2$, and $ V_n^{(p)}(t)$ displays a monotonic step-like increase. In the right panel shown is the test for $p = 1/H$ for $H = 0.1$, and as expected $V_n^{(p)}(t)$ tends to zero as $n$ is increased. }
	\label{fig:pvar_stork1}
\end{figure}

\begin{figure}[h!]
	\center
	\includegraphics[width=.9\linewidth]{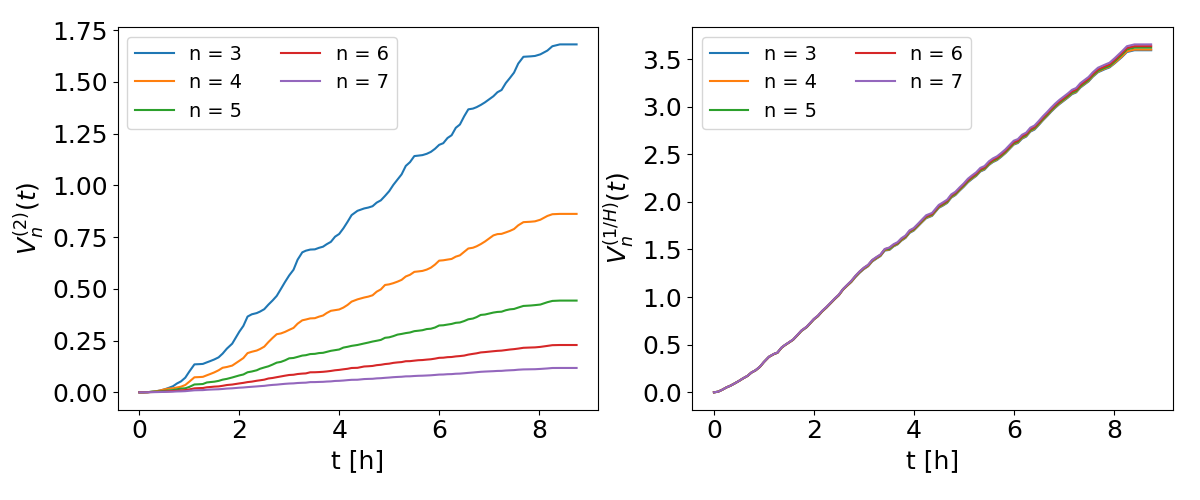}
	\caption{A p-variation test on a randomly chosen daily segment of the migrating stork (September). In the left panel $p = 2$, and $ V_n^{(p)}(t)$, while in the right panel shown is the test for $p = 1/H$ for $H = 1$. Both agree with the theoretical prediction for FBM-like dynamics.  }
	\label{fig:pvar_stork3}
\end{figure}

Here we corroborate several of the claims made in the main text, by applying a p-variation test to randomly chosen trajectories, see Appendix E in the main text. The test is applied to amoeba (Fig.~\ref{fig:pvar_amoeba}), kite (Fig.~\ref{fig:pvar_kite}) and stork (Figs.~\ref{fig:pvar_stork1}-\ref{fig:pvar_stork3}) trajectories.

\end{document}